\documentclass[iop, twocolappendix]{mn2e} 

\usepackage[pdftex]{graphicx}

\usepackage{amsmath, amsthm, amssymb}
\let\bibhang\relax
\usepackage{natbib}
\usepackage{subfigure}
\usepackage{upgreek}
\usepackage{accents}
\usepackage{color}
\usepackage{hyperref}
\usepackage{dblfloatfix}
\usepackage[T1]{fontenc}
\usepackage{aecompl}

\newcommand*{\dt}[1]{%
  \accentset{\mbox{\large\bfseries .}}{#1}}

\newcommand{\emcee}{\textsc{emcee}}

\newcommand{\steep}{\texttt{steep}}
\newcommand{\dpl}{\texttt{dpl}}
\newcommand{\floor}{\texttt{floor}}

\newcommand{\Omnow}{\Omega_{\text{m},0}}
\newcommand{\Obnow}{\Omega_{\text{b},0}}

\newcommand{\Tcmb}{T_{\gamma}}

\newcommand{\TS}{T_{\text{S}}}
\newcommand{\TK}{T_{\text{K}}}

\newcommand{\heat}{\text{heat}}

\newcommand{\Lya}{\text{Ly-}\alpha}
\newcommand{\Lyn}{\text{Ly-}n}

\newcommand{\LyC}{\text{LyC}}
\newcommand{\LW}{\text{LW}}

\newcommand{\NHI}{N_{\text{H } \textsc{i}}}
\newcommand{\NHeI}{N_{\text{He } \textsc{i}}}

\newcommand{\gammalo}{\gamma_{\text{lo}}}
\newcommand{\gammahi}{\gamma_{\text{hi}}}

\newcommand{\MUV}{M_{\text{UV}}}

\newcommand{\eheat}{\upepsilon_{\heat}}

\newcommand{\enu}{\upepsilon_{\nu}}

\newcommand{\xibar}{\overline{x}_i}

\newcommand{\zrei}{z_{\text{rei}}}

\newcommand{\Mbh}{M_{\bullet}}

\newcommand{\fstar}{f_{\ast}}

\newcommand{\rhostardot}{\dt{\rho}_{\ast}}

\newcommand{\Nion}{N_{\text{ion}}}
\newcommand{\Nlw}{N_{\text{LW}}}

\newcommand{\fesc}{f_{\text{esc}}}
\newcommand{\fescnu}{f_{\text{esc},\nu}}
\newcommand{\fescLyC}{f_{\text{esc}, \text{LyC}}}
\newcommand{\fescLW}{f_{\text{esc}, \text{LW}}}
\newcommand{\Msun}{M_{\odot}}

\newcommand{\Tmin}{T_{\text{min}}}

\newcommand{\sigmaHI}{\sigma_{\text{H} {\textsc{i}},\nu}}
\newcommand{\sigmaHeI}{\sigma_{\text{He} {\textsc{i}},\nu}}

\newcommand{\fheat}{f_{\text{heat}}}

\newcommand{\dTb}{\delta T_b}

\newcommand{\intensityunitsnumber}{\text{s}^{-1} \ \text{cm}^{-2} \ \mathrm{Hz}^{-1} \ \text{sr}^{-1}}

\newcommand{\coheatingdensity}{\text{erg} \ \text{s}^{-1} \ \text{cMpc}^{-3}}

\newcommand{\cXunits}{\text{erg} \ \text{s}^{-1} \ (\Msun \ \text{yr})^{-1}}
\newcommand{\sfrdunits}{\Msun \ \text{yr}^{-1} \ \text{cMpc}^{-3}}

\newcommand{\QHII}{Q_{\textsc{HII}}}

\title[The global 21-cm signal in context]{The global 21-cm signal in the context of the high-z galaxy luminosity function}
\author[Mirocha, Furlanetto, \& Sun]{
Jordan Mirocha$^1$\textsuperscript{\thanks{mirocha@astro.ucla.edu}}, Steven R. Furlanetto$^1$, \& Guochao Sun$^2$
\\
$^{1}$Department of Physics and Astronomy, University of California, Los Angeles, CA 90024, USA\\
$^{2}$Cahill Center for Astronomy and Astrophysics, California Institute of Technology, Pasadena, CA 91125, USA
}

\begin{document}

\pagerange{\pageref{firstpage}--\pageref{lastpage}} \pubyear{2016}
\maketitle

\begin{abstract}
We build a new model for the global 21-cm signal that is calibrated to measurements of the high-$z$ galaxy luminosity function (LF) and further tuned to match the Thomson scattering optical depth of the cosmic microwave background, $\tau_e$. Assuming that the $z \lesssim 8$ galaxy population can be smoothly extrapolated to higher redshifts, the recent decline in best-fit values of $\tau_e$ and the inefficient heating induced by X-ray binaries (the presumptive sources of the high-$z$ X-ray background) imply that the entirety of cosmic reionization and reheating occurs at $z \lesssim 12$. In contrast to past global 21-cm models, whose $z \sim 20$ ($\nu \sim 70$ MHz) absorption features and strong $\sim 25$ mK emission features were driven largely by the assumption of efficient early star-formation and X-ray heating, our new models peak in absorption at $\nu \sim 110$ MHz at depths $\sim -160$ mK and have negligible emission components. Current uncertainties in the faint-end of the LF, binary populations in star-forming galaxies, and UV and X-ray escape fractions introduce $\sim 20$ MHz ($\sim 50$ mK) deviations in the trough's frequency (amplitude), while emission signals remain weak ($\lesssim 10$ mK) and are confined to $\nu \gtrsim 140$ MHz. These predictions, which are intentionally conservative, suggest that the detection of a 21-cm absorption minimum at frequencies below $\sim 90$ MHz and/or emission signals stronger than $\sim 10$ mK at $\nu \lesssim 140$ MHz would provide strong evidence for ``new'' sources at high redshifts, such as Population III stars and their remnants.
\end{abstract}

\section{Introduction} \label{sec:Introduction}
Galaxy evolution in the early Universe is most often studied from two distinct vantage points: through direct observations of high-$z$ galaxies and measurements of the thermal and ionization history of the intergalactic medium (IGM). These two approaches are exceptionally complementary, at least in principle, as the IGM is a repository of photons that never reach our telescopes. At the highest redshifts, due to limitations of even the most powerful space-based optical and near-infrared instrumentation, future constraints on the properties of the IGM will serve as an essential substitute for direct observations of galaxies themselves. As a result, the establishment of a framework for inferring galaxy properties from IGM signals is paramount.

The canonical probe of high-$z$ galaxies is the rest-frame ultraviolet (UV) galaxy luminosity function (LF), i.e., the number density of galaxies per unit luminosity and redshift. Dedicated programmes using the \textit{Hubble Space Telescope} (\textit{HST}) have driven progress in this area at the highest redshifts so far probed, with healthy samples now extending to redshifts as high as $z \sim 8$ \citep{Bouwens2015,Finkelstein2015}, and a number of candidates at $9 \lesssim z \lesssim 12$ \citep[e.g.,][]{Ellis2013, Oesch2014}. The \textit{James Webb Space Telescope} (\textit{JWST}) will be very important for filling out the sample of galaxies at yet higher redshifts, but current models predict its reach will not extend beyond $z \sim 15$ \citep[e.g.,][]{Mason2015} without the aid of strong lensing, which has boosted \textit{HST}'s capabilities in the Frontier Fields \citep{Atek2015,Livermore2016}.

Complementary IGM-based constraints on high-$z$ galaxies are far more crude at this stage. The Thomson Scattering optical depth, $\tau_e$, for example, constrains the total column density of electrons between the observer and the cosmic microwave background (CMB), while Gunn-Peterson troughs in quasar spectra mark the end of reionization at $z \sim 6$ \citep[e.g.,][]{Fan2002}. Together, these constraints provide a lower limit on the duration of the Epoch of Reionization (EoR), which one can parameterize as a reionization redshift, $z_{\mathrm{rei}}$, assuming an instantaneous transition from neutral to ionized. The \textit{Planck} team recently reported a Thomson scattering optical depth to the cosmic microwave background (CMB) of $\tau_e = 0.055 \pm 0.009$ \citep{Planck2016}, indicating $z_{\mathrm{rei}} \sim 8 \pm 1$, and thus a minimal duration of $\Delta \zrei \sim 2$. 

Efforts to jointly interpret the aforementioned measurements have largely been geared toward reconciliation. Do the number of photons generated by the galaxies we do see match the number required to maintain a state of full ionization in the IGM? Furthermore, is the population of galaxies in place prior to full reionization substantial enough to match the most up-to-date measurements of $\tau_e$? The answer to both of these questions is ``yes,'' provided one makes reasonable assumptions about (i) the abundance of galaxies we do \textit{not} see (i.e., extrapolations to the LF), and (ii) the escape fraction of Lyman continuum (LyC) photons from galaxies, $\fescLyC$. Recent work suggests that extrapolating the Schechter form of the LF to low luminosities may well be reasonable \citep{Livermore2016}, at least at $z \sim 6$, while the $\fescLyC \sim 0.2$ values that have caused discomfort in recent years may be now reasonable if the UV emission of star clusters is boosted by binary star evolution \citep{Eldridge2009,Stanway2016,Ma2016}.

Despite such reduced tensions between theory and observation, the story of galaxy evolution in the early Universe is far from complete. In the coming years, observations of redshifted 21-cm emission from neutral hydrogen are expected to contribute substantially to our existing understanding of reionization and high-$z$ galaxies while opening up a brand new window into the excitation (or ``spin'') temperature history of the high-$z$ IGM \citep[e.g.,][]{Madau1997,FurlanettoOhBriggs2006}. As a result, 21-cm measurements promise to weigh in on long-standing questions regarding the ionizing photon production efficiency of high-$z$ galaxies, the nature of X-ray sources \citep[e.g.,]{Pritchard2007,Fialkov2014,Pacucci2014,EwallWice2016}, and perhaps even the properties of the interstellar medium of high-$z$ galaxies, which can serve as both a source of radiation \citep[e.g., bremsstrahlung;][]{Mineo2012b} and sink for LyC, X-ray, and perhaps even Lyman-Werner photons \citep{Kitayama2004,Schauer2015}, whose escape fraction we consider in Section \ref{sec:escape} for the first time in a 21-cm context.

Studies aimed at better-understanding the complementarity of 21-cm measurements and other EoR probes, though few so far, demonstrate that even crude 21-cm constraints can greatly aid our understanding of reionization and high-$z$ galaxies \citep{Pritchard2010b, Beardsley2015}. Preliminary results from the \textit{Precision Array for Probing the Epoch of Reionization} (\textit{PAPER}) have since bolstered these arguments, finding the first observational evidence of X-ray heating of the high-$z$ IGM through upper limits on the 21-cm power spectrum \citep{Parsons2014,Ali2015,Pober2015,Greig2016}, and thus constrained the X-ray properties of $z \sim 8$ galaxies for the first time. The complementarity can be viewed from the opposite perspective as well, since constraints on high-$z$ galaxies can in principle be used to better separate signal from foreground \citep{Petrovic2011}.

The sky-averaged (``global'') 21-cm signal \citep{Shaver1999}, now being targeted by several ground-based experiments \citep[e.g., EDGES, BIGHORNS, SCI-HI, SARAS, LEDA;][]{Bowman2010,Sokolowski2015,Voytek2014,Patra2015,Bernardi2016}, with more concepts in design \citep[e.g., DARE;][]{Burns2012}, is a particularly clear-cut ally of galaxy surveys as it is sensitive to the volume-averaged (i.e., luminosity function integrated) emissivity of galaxies. The mean ionization and spin temperature histories encoded by the global 21-cm signal of course influence the 21-cm power spectrum as well. The joint constraining power of the power spectrum and global signal was recently considered by \citet{Liu2016b}, though to the best of our knowledge the 21-cm signal (sky average or power spectrum) and galaxy LF have yet to be considered in a common framework. This has prevented 21-cm models from calibrating to recent advances driven by \textit{HST}, and as a result has led to predictions spanning the a wide range of possibilities \citep[e.g.,][]{Furlanetto2006,Pritchard2010a,Mesinger2013,Fialkov2014,Mirocha2015,Tanaka2016}. 

Our goal here is to address these issues in two steps:
\begin{enumerate}
    \item Leverage the success of simple models for the galaxy LF \citep[e.g.,][]{Trenti2010,Tacchella2013,Sun2016} to create a new ``vanilla model'' for the global 21-cm signal calibrated both to the LF \citep{Bouwens2015} and $\tau_e$ \citep{Planck2016}. 
    \item Explore simple extensions to the standard picture of the LF in an attempt to determine the global 21-cm signal's sensitivity to the properties of the faint galaxy population, and thus more concretely determine how its detection will complement future galaxy surveys and 21-cm power spectrum experiments. 
\end{enumerate}
In the near term, these models can be used to test signal extraction algorithms and better inform instrument design. In the longer term, our models will provide a reference point from which to interpret a global 21-cm measurement in the broader context of galaxy formation. 

This paper is organized as follows. In Section \ref{sec:model}, we outline our theoretical model for the galaxy population and global 21-cm signal. In Section \ref{sec:results}, we present our main results, including our LF-calibrated model for the global 21-cm signal, its sensitivity to the star-formation efficiency of faint galaxies, the stellar and black hole populations of high-$z$ galaxies, and the escape fraction of UV and X-ray photons. We provide some discussion of our results in Section \ref{sec:discussion} and summarize our main conclusions in Section \ref{sec:conclusions}. We use cosmological parameters from \citet{Planck2015} throughout.

\section{Modeling Framework} \label{sec:model}
Our model has essentially two parts: (i) a model for the galaxy population, whose properties are derived from the dark matter halo mass function and assumptions about the relationship between halo mass and halo luminosity, and (ii) a model for the global 21-cm signal, which takes the volume-averaged emissivity of the galaxy population as input, and from it determines the ionization and thermal history of the IGM. Though its individual components are similar to models appearing in the literature in recent years, in this section we briefly outline the procedure to emphasize the connection between the galaxy LF and global 21-cm signal, which to our knowledge have yet to be considered in a common framework.

\subsection{Constructing the UV Luminosity Function} \label{sec:LF}
Our model is motivated by recent studies of the high-$z$ galaxy LF based on abundance matching \citep{Mason2015,Mashian2016,Sun2016}. The mismatch between the shape of the dark matter halo mass function (HMF) and the galaxy LF can be accounted for by (i) a mass-dependent occupation fraction or duty cycle, $f_{\mathrm{DC}}$, of galaxies in halos \citep[e.g.,][]{Trenti2010} and/or (ii) differential evolution of galaxy luminosity, $L_h$, with halo mass, $M_h$, and/or redshift, $z$ \citep[e.g.,][]{Mason2015,Mashian2016,Sun2016}. 

Unfortunately, these two approaches cannot be distinguished by measurements of luminosity functions alone. Because there is theoretical support for consistent active star-formation in halos at high-$z$ \citep[except perhaps in very low-mass halos; e.g,][]{OShea2015,Xu2016}, we will operate within the $L_h = L_h(M_h, z)$ framework (scenario (i)) rather than invoking $f_{\mathrm{DC}} < 1$.

In this case, the \textit{intrinsic} luminosity function can be expressed as
\begin{equation}
    d\phi(L_h) = \frac{d n(M_h, z)}{dM_h} \left( \frac{dL_h}{d M_h} \right)^{-1} d L_h . \label{eq:LF}
\end{equation}
where $n(M_h, z)$ is the number density of halos of mass $M_h$ at redshift $z$, and $\phi$ is the galaxy LF. Equation \ref{eq:LF} is equivalent to the \textit{observed} LF only under the assumption that all photons in the observed band (rest-frame $1600$\AA\ here) escape galaxies. In general, this is not the case, as some rest-frame $1600$\AA\ photons will be absorbed by dust before they can escape the galaxy. However, we will ignore dust in what follows as our calculations are restricted to $z \gtrsim 6$, at which time dust extinction has only a minor impact on the conversion between the observed and intrinsic LF \citep[e.g., ][]{Bouwens2012,Capak2015}.

Recent work suggests that $L_h$ must evolve with redshift if it is to fit all  high-$z$ data \citep[e.g.,][]{Mason2015,Mashian2016,Sun2016}. One approach is to parameterize $L_h$ directly as a function of both halo mass and redshift and solve for the dependencies required to fit high-$z$ LFs \citep{Mashian2016}. We will adopt a slightly different approach, which we describe below.

First, because the observed LF of high-$z$ galaxies probes rest-frame UV luminosity, which is dominated by massive young stars, we take the \textit{intrinsic} luminosity of galaxies to be
\begin{equation}
    L_{h,\nu} = \dot{M}_{\ast} (M_h, z) \mathcal{L}_{\nu}  \label{eq:Lh}
\end{equation}
where $\dot{M}_{\ast}$ is the star-formation rate (SFR) and $\mathcal{L}_{\nu}$ is a conversion factor which sets the luminosity (in band $\nu$) per unit star-formation. 

In general, $\mathcal{L}_{\nu}$, could depend on $M_h$ and $z$, though for the remainder of this study we will assume it is a constant. For our fiducial models we adopt the \textsc{BPASS} version 1.0 models \textit{without} binaries \citep{Eldridge2009}, from which the photon production at $1600$\AA, $\mathcal{L}_{1600}$, and in the LW and LyC bands, $\mathcal{L}_{\LyC}$ and $\mathcal{L}_{\LW}$, can be computed self-consistently from a choice of the stellar metallicity, $Z$, and in general the stellar initial mass function, nebular emission, and so on. We compare results obtained using models \textit{with} binaries, as well as the those obtained using \textsc{starburst99} \citep{Leitherer1999} in Appendix \ref{appendix:spop}.

We model the SFR as \citep[e.g.,][]{Sun2016}
\begin{equation}
    \dot{M}_{\ast}(M_h, z) = \fstar(M_h, z) \dot{M}_b(z, M_h) \label{eq:SFR}
\end{equation}
where $\dot{M}_b(z, M_h)$ is the baryonic mass accretion rate (MAR) onto a dark matter (DM) halo of mass $M_h$ at redshift $z$, and $\fstar$ is the efficiency of star formation. The baryonic MAR is well approximated by \citep[e.g.,][]{McBride2009,Dekel2013}
\begin{equation}
    \dot{M}_b(z, M_h) = f_b \dot{M}_{h,0} M_h^{\gamma_M} (1 + z)^{\gamma_z} \label{eq:bMAR}
\end{equation}
with $\gamma_M \sim 1$ and $\gamma_z \sim 5/2$, where $\dot{M}_{h,0}$ is a normalization constant and $f_b$ is the primordial baryon fraction.

However, rather than adopting a parametric form for the MAR calibrated by simulations, we derive it directly from the halo mass function. The basic idea is to abundance match halos across redshifts, and in so doing determine their trajectories through mass space. This approach ensures self-consistency, as, for example, a population of halos evolved forward in time through an MAR model will not necessarily match an independently-generated model for the halo mass function at all times. See Furlanetto et al., in prep. for more details.

For our fiducial model, we assume the star formation efficiency (SFE) is a double power-law (DPL) in $M_h$, i.e., 
\begin{equation}
    \fstar(M_h) = \frac{f_{\ast,0}} {\left(\frac{M_h}{M_{\mathrm{p}}} \right)^{\gamma_{\mathrm{lo}}} + \left(\frac{M_h}{M_{\mathrm{p}}} \right)^{\gamma_{\mathrm{hi}}}} \label{eq:sfe_dpl}
\end{equation}
where $f_{\ast,0}$ is the SFE at its peak mass, $M_p$, and $\gammalo$ and $\gammahi$ describe the power-law index at low and high masses, respectively. We will consider two additional modifications to this form in \S\ref{sec:results}, and the possibility of a redshift dependence in \S\ref{sec:discussion}.

Equations \ref{eq:LF}-\ref{eq:sfe_dpl} define our model for the galaxy LF, which is a hybrid between physical and empirical models, as the foundation of our model is the HMF but we treat the SFE using a phenomenological model whose parameters must be calibrated by observations. As a result, we will need to extrapolate the SFE to lower mass halos and higher redshifts than are represented in our calibration dataset in order to model the global 21-cm signal. Part of our goal in Section \ref{sec:results} is to determine how these extrapolations to the SFE -- which are equivalent to extrapolations in the galaxy LF -- affect the global 21-cm signal.

\subsection{Generating the Global 21-cm Signal}
We adopt the commonly-used two-zone model for the the IGM \citep[e.g.,][]{Furlanetto2006,Pritchard2010a} in which the volume-filling factor of ionized regions ($\QHII$) and the electron fraction ($x_e$) and kinetic temperature ($\TK$) of the ``bulk IGM'' outside fully ionized regions are treated separately. In order to solve for these three quantities, one requires (i) a model for the volume-averaged emissivity of galaxies, $\enu(z)$, (ii) an algorithm to compute the mean background intensity, $J_{\nu}$, i.e., the volume-averaged emissivity attenuated by neutral IGM gas and diluted by cosmic expansion, and (iii) a non-equilibrium chemical reaction network which solves for the evolution of $\QHII$, $x_e$, and $T_K$ in response to the photo-ionization and heating rates set by $J_{\nu}$. 

With a model for the luminosities of individual galaxies, the volume-averaged emissivity can be computed via integration of the galaxy LF weighted by the escape fraction and luminosity in the relevant band, 
\begin{equation} 
    \enu(z) = \int_{L_{\min}}^{\infty} \fescnu L_{h,\nu} \frac{d \phi(L_{h,{\nu}})}{dL_{h,{\nu}}} dL_{h,{\nu}} \label{eq:emissivity} 
\end{equation}
We allow the escape fraction in the 10.2-13.6 eV band (which we refer to loosely as the Lyman-Werner (LW) band), $\fescLW$, to differ from the Lyman-continuum (LyC) escape fraction, $\fescLyC$, though both are treated as constant, frequency-independent quantities. For X-rays, it is more sensible to parameterize the escape fraction with a neutral hydrogen column density, which hardens the intrinsic X-ray spectrum differentially as a function of frequency, i.e.,
\begin{equation}
    \fescnu = \exp\left[-\NHI (\sigmaHI + y \sigmaHeI) \right] . \label{eq:fescX}
\end{equation}
Here, $y$ is the primordial helium abundance by number, and $\sigmaHI$ and $\sigmaHeI$ are the bound-free cross sections for neutral hydrogen and neutral helium, respectively. Note that this expression assumes that $\NHeI = y \NHI$, i.e., the neutral helium fraction is equal to the neutral hydrogen fraction, and ignores the opacity of singly ionized helium.

With models for the volume-averaged emissivity in hand, the ionization and heating rates in each phase of the IGM can be computed, and the ionization states and kinetic temperatures evolved forward in time. For the fully ionized phase of the IGM, the ionization rate governs the rate at which the volume of ionized bubbles grows, while the temperature is held fixed at $\TK = 2 \times 10^4$ K. 

Solving for the ionization state and temperature of the bulk IGM is more challenging. We solve the cosmological radiative transfer equation to determine the meta-galactic radiation background intensity, $J_{\nu}$, which can be done efficiently assuming a neutral high-$z$ IGM\footnote{Accounting for the late-time softening of the meta-galactic X-ray background as reionization progresses has a negligible impact on our results.} \citep{Mirocha2014}. 

We model $\mathcal{L}_{\nu}$ for X-ray sources as a multi-colour disk (MCD) spectrum \citep{Mitsuda1984} appropriate for high-mass X-ray binaries (HMXBs), which are now known to be important for reionization \citep{Power2013,Fragos2013} and 21-cm models \citep[e.g.,][]{Fialkov2014,Mirocha2014}. We assume black holes with masses $\Mbh = 10 \Msun$ and normalize the MCD spectrum to the observed X-ray luminosity star-formation rate ($L_X$-SFR) relation, which we will discuss in more detail in \S\ref{sec:spop}.

The differential brightness temperature can then be computed as \citep[e.g.,][]{Furlanetto2006}
\begin{equation}
    \dTb \simeq 27 (1 - \xibar) \left(\frac{\Obnow h^2}{0.023} \right) \left(\frac{0.15}{\Omnow h^2} \frac{1 + z}{10} \right)^{1/2} \left(1 - \frac{\Tcmb}{T_{\mathrm{S}}} \right) , \label{eq:dTb}
\end{equation}
where $\xibar = \QHII + (1 - \QHII) x_e$ and
\begin{equation}
    T_S^{-1} \approx \frac{T_{\gamma}^{-1} + x_c T_K^{-1} + x_{\alpha} T_{\alpha}^{-1}}{1 + x_c + x_{\alpha}}
\end{equation}
is the excitation or ``spin'' temperature of neutral hydrogen, which characterizes the number of hydrogen atoms in the hyperfine triplet state relative to the singlet state, and $T_{\alpha} \simeq T_K$. We compute the collisional coupling coefficient, $x_c$, using the tabulated values in \citet{Zygelman2005} and take the radiative coupling coefficient \citep{Wouthuysen1952,Field1958} to be $x_{\alpha} = 1.81 \times 10^{11} \widehat{J}_{\alpha} S_{\alpha} / (1 + z)$, where $S_{\alpha}$ is a factor of order unity that accounts for line profile effects \citep{Chen2004,FurlanettoPritchard2006,Chuzhoy2006,Hirata2006}, and $\widehat{J}_{\alpha}$ is the intensity of the $\Lya$ background in units of $\intensityunitsnumber$.
 
All calculations were carried out with the \textsc{ares} code\footnote{https://bitbucket.org/mirochaj/ares}. Within \textsc{ares}, we use the \textsc{hmf-calc} code \citep{Murray2013}, which itself depends on the \textit{Code for Anisotropies in the Microwave Background} \citep[\textsc{camb};][]{Lewis2000}, to compute the HMF, which we take to be the \citet{ShethMoTormen2001} form. Regarding the Wouthuysen-Field coupling, we use the analytic formulae from \citet{FurlanettoPritchard2006} to determine $S_{\alpha}$ and adopt the recycling fractions (of $\Lyn \rightarrow \Lya$ photons) from \citep{Pritchard2006}. We use the lookup tables of \citet{Furlanetto2010} to determine the fraction of photo-electron energy deposited as heat and further ionization, bound-free absorption cross-sections  of \citet{Verner1996}, while recombination and cooling rate coefficients were taken from \citet{Fukugita1994}. We generate initial conditions using the \textsc{CosmoRec} code \citep{Chluba2011}, and adopt \textit{Planck} cosmological parameters \citep{Planck2015}. Fiducial parameters are listed in Table \ref{tab:parameters}.

\begin{table}
\begin{tabular}{ | l | l | l }
\hline
name & value & description \\
\hline
$\fescLW$   & 1.0 & escape fraction of 10.2-13.6 eV photons \\
$\fescLyC$  & 0.1 & escape fraction of LyC photons \\
$\NHI$      & 0   & absorbing column of X-ray sources \\
$\Tmin$ /K  & $10^4$ & virial temperature threshold \\
$Z$         & $0.02$ & stellar metallicity \\
$\fstar$    & \dpl & functional form of SFE \\
\hline
$f_{\ast,0}$ & 0.05  & SFE normalization \\
$M_p/\Msun$        & $2.8\times 10^{11}$  & halo mass at which $\fstar$ peaks \\
$\gammalo$   & 0.49   & low-mass slope of $\fstar(M_h)$\\
$\gammahi$   & -0.61  & high-mass slope of $\fstar(M_h)$
\end{tabular}
\caption{Fiducial model parameters. Note that the last four rows contain the best-fit values for the parameters of Equation \ref{eq:sfe_dpl}, which are obtained through our calibration procedure (see \S\ref{sec:calibration}).}
\label{tab:parameters}
\end{table}

\section{Results} \label{sec:results}
In this section we present first the calibrated SFE model and its uncertainties (\S\ref{sec:calibration}) along with the resultant LF and global 21-cm signal predictions (\S\ref{sec:vanilla}). Then, we proceed to investigate how sensitive the global 21-cm signal is to the minimum mass of star-forming galaxies (\S\ref{sec:sfe}), their stellar and black hole populations (\S\ref{sec:spop}), and the escape fraction of Lyman-Werner, Lyman continuum, and X-ray photons (\S\ref{sec:escape}). Because we are focused principally on ``vanilla'' models, i.e., realizations of the signal brought about by  galaxies as we (think we) understand them, we do not in this section consider contributions from PopIII stars, ``miniquasars,'' or other non-standard or unobserved sources of the high-$z$ LW, LyC, and X-ray radiation backgrounds. We will discuss such sources again in Section \ref{sec:discussion}.

\subsection{Model Calibration} \label{sec:calibration}
We fit for the four parameters of Equation \ref{eq:sfe_dpl} using the affine-invariant Markov Chain Monte Carlo code \emcee\ \citep{ForemanMackey2013}. We adopt the \citet{Bouwens2015} measurements of the luminosity function at $z\sim 6,7,$ and 8 as our calibration dataset (their Table 5). Our likelihood function is then
\begin{equation}
    l(\mathbf{x} | \mathbf{\theta}) = \Pi_i^{N_z} \Pi_j^{N_M} p_{ij}
\end{equation}
where the index $i$ runs over redshift bins and $j$ runs over magnitude bins. $p_{ij}$ is the probability of the data, $\mathbf{x}$, given the model, described by parameters $\mathbf{\theta}$, i.e.,
\begin{equation}
    p_{ij} = \frac{1}{\sqrt{2\pi \sigma_{ij}^2}} \exp\bigg\{-\frac{[\phi(z_i, M_j) - \phi(z_i, M_j | \mathbf{\theta})]^2}{2 \sigma_{ij}^2} \bigg\}
\end{equation}
Here, $\sigma_{ij}$ is the uncertainty in the number density of galaxies at redshift $z_i$ in magnitude bin $M_j$, which we force to be Gaussian when necessary simply by averaging the occasional asymmetric error bar quoted in \citet{Bouwens2015}. We adopt broad uninformative priors on each model parameter of interest. 

\begin{figure}
\begin{center}
\includegraphics[width=0.48\textwidth]{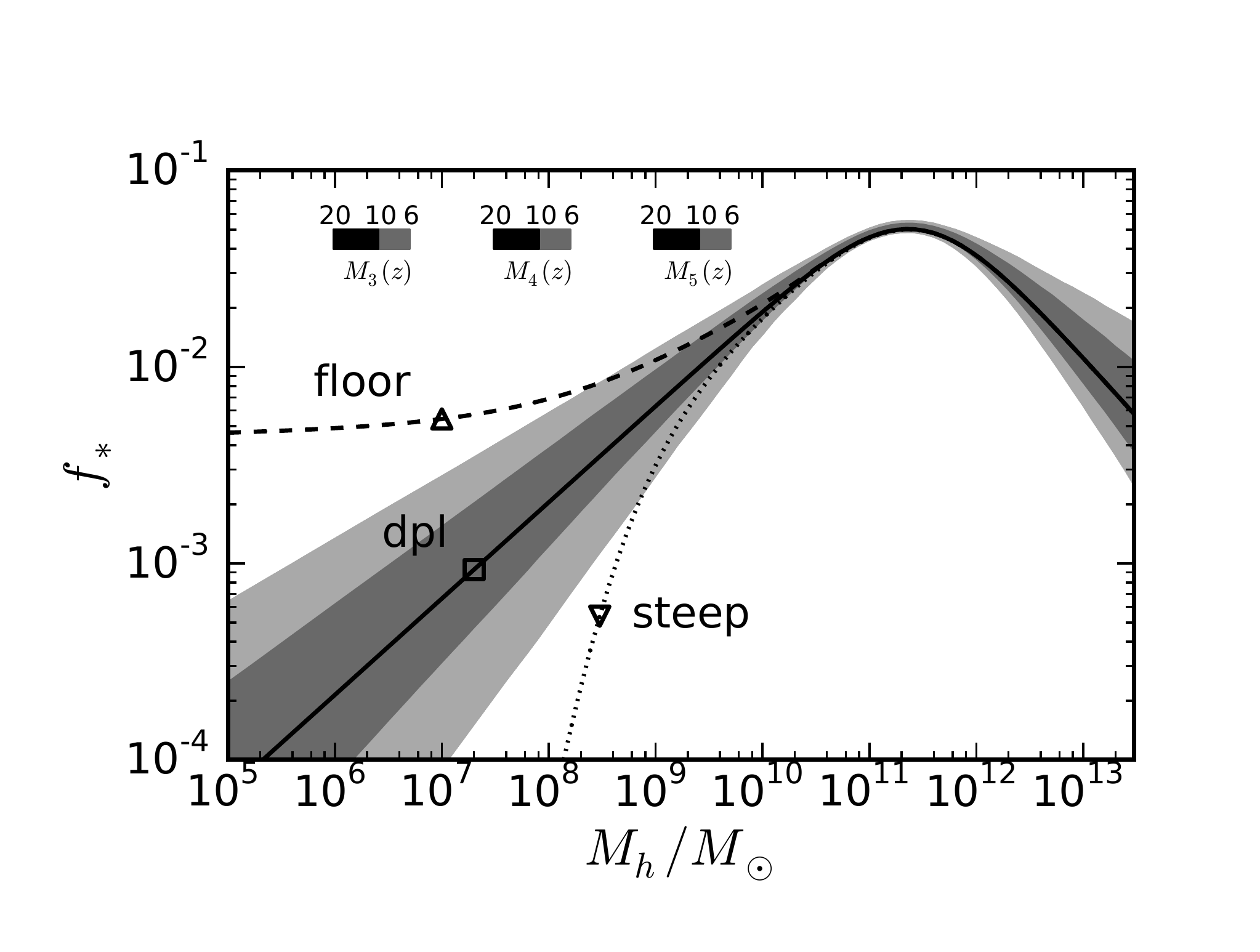}
\caption{Calibrated SFE curve computed using the \citet{Bouwens2015} data at $z \sim 6$. The solid black line is the maximum likelihood SFE curve, with dark and light grey regions denoting regions containing 68\% and 95\% of the likelihood, respectively. Note that the boundaries of shaded regions do not necessarily correspond to realizations of the SFE curve, as they are determined through computation of the 68\% and 95\% ranges for each $M_h$ independently. Shown for reference are the halo masses corresponding to virial temperatures of $10^3$, $10^4$, and $10^5$ K at redshifts $z=6,10,$ and 20 (upper left). The dashed and dotted curves are simple extensions to the pure double power-law SFE curve explored in subsequent figures. $\MUV = -17$ corresponds to a halo mass of $\sim 10^9 M_{\odot}$ for our fiducial model.}
\label{fig:calib_sfe}
\end{center}
\end{figure}

In Figure \ref{fig:calib_sfe} we show our reconstructed SFE curve (solid black line) and its 68\% and 95\% confidence intervals (dark and light grey regions, respectively), both of which agree well with other recent work \citep{Mason2015,Sun2016}. Note the characteristic peak at $M_h \sim 3 \times 10^{11} \Msun$ and the peak efficiency of $\sim 0.05$, in contrast to the common assumption of a constant ``effective'' efficiency in all halos above the atomic cooling threshold. The overall normalization of the SFE is uncertain by a factor of several (not pictured in Figure \ref{fig:calib_sfe}), as it depends on the assumed stellar population, halo mass function, and in general, the dust correction. However, this uncertainty in the normalization does \textit{not} translate to a comparable uncertainty in the volume-averaged ionization or thermal history, since the ionization and heating rates depend on the luminosity \textit{density}. We will revisit this point in Section \ref{sec:spop} and Appendix \ref{appendix:spop}.

The uncertainties in $\fstar$ grow at both the low- and high-mass ends. Though the SFE of high-mass galaxies almost certainly encodes interesting physics,  such sources are probably too rare to make an impact on the global 21-cm signal. As a result, we will focus principally on possible behaviors in the low-mass end of the SFE, equivalent to the faint-end of the galaxy LF, and leave questions regarding the bright-end to be addressed properly by \textit{WFIRST} in the years to come.

We will consider two simple phenomenological extensions to the pure double power law SFE model in the remainder of the paper, which we refer to as the \floor\ and \steep\ models, as identified by dashed and dotted lines in Figure \ref{fig:calib_sfe}, respectively. The \floor\ model is implemented by adding a constant 0.5\% SFE in halos below $10^9 \ \Msun$, while the \steep\ model is a multiplicative modification to the \dpl\ model of the form
\begin{equation}
    f_{\ast}(M_h) = \bigg[1 + \left(2^{\mu / 3)} - 1\right) \left(\frac{M_h}{M_c} \right)^{-\mu} \bigg]^{-3 / \mu} ,
\end{equation}    
with $\mu = 1$ and $M_c = 10^{10} \Msun$. The \floor\ model resembles some physical models \citep[e.g.][]{Mason2015}, while the \steep\ model is similar (though much more extreme) than occupation fraction predictions from recent numerical simulations \citep[e.g.,][who found $\mu \sim 1.5$ and $M_c \sim 6 \times 10^7 \Msun$]{OShea2015}. However, our main motivation for adopting these particular extensions to the SFE is that their corresponding $\tau_e$ values differ by an amount comparable to the $1-\sigma$ confidence interval for $\tau_e$ recently published by the \textit{Planck} team, assuming $\fesc = 0.1$ and $\Tmin = 10^4$ K, as we will see in the next section.

\subsection{The Luminosity Function -- Global 21-cm Connection} \label{sec:vanilla}
\begin{figure*}
\begin{center}
\includegraphics[width=0.98\textwidth]{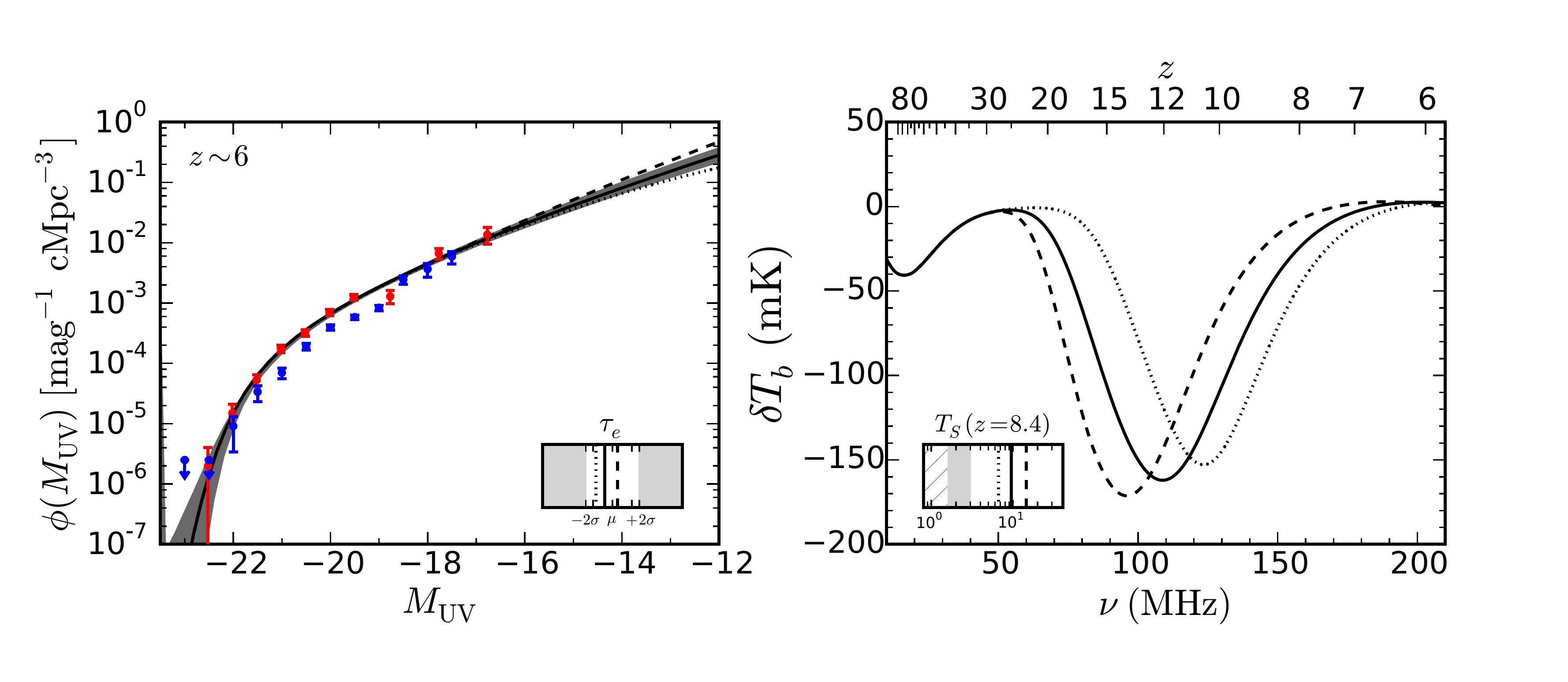}
\caption{\textit{Left:} Luminosity functions at $z \sim 6$ using the SFE curves shown in Figure \ref{fig:calib_sfe}. Observational data from \citet{Bouwens2015} and \citet{Finkelstein2015} are shown in red and blue, respectively, though only the red points were used in the fit. The grey region is the 68\% confidence interval, and dashed and dotted curves are the same extensions to the fiducial \dpl\ SFE model as shown in Figure \ref{fig:calib_sfe}. Inset shows the CMB optical depth, relative to the recent \textit{Planck} measurement of $\tau_e = \mu \pm \sigma$, with $(\mu,\sigma) = (0.055, 0.009)$. \textit{Right:} Models of the global 21-cm signal corresponding to the LF models of the left panel. The inset shows the mean IGM spin temperature of each model at $z=8.4$ relative to the unphysical adiabatic cooling limit (hatched region) and lower limits from \textit{PAPER} (shaded region).}
\label{fig:gs_fiducial}
\end{center}
\end{figure*}

In Figure \ref{fig:gs_fiducial}, we present the main result of this paper: models for the galaxy LF and corresponding predictions for the global 21-cm signal. We match the $z\sim 6$ LF as measured by \citet{Bouwens2015} by construction. Our best fit model is again represented by the solid black line, with 68\% confidence region denoted by the shaded grey region, and phenomenological SFE \floor\ and \steep\ extensions shown as dashed and dotted curves. The inset shows the Thomson optical depth of the CMB for each SFE model, relative to the \textit{Planck} measurement of $\tau_e = 0.055 \pm 0.009$ \citep{Planck2016}, assuming $\fescLyC = 0.1$. 

The SFE behavior affects the predicted global 21-cm signal as well, as shown in the right panel of Figure \ref{fig:gs_fiducial}. Imposing a \floor\ in the SFE in low-mass galaxies causes earlier features in the global 21-cm signal, while a \steep\ decline in the SFE causes later features. All three curves are deep ($\sim -160 \pm 10$ mK), and occur at frequencies $95 \lesssim \nu \lesssim 125$ MHz. The \dpl\ and \floor\ models have a very weak $\lesssim 5$ mK emission feature, which would be extremely difficult to detect observationally. Indeed, the \steep\ model lacks an emission feature entirely. 

Realizations of the global 21-cm signal similar to those of Figure \ref{fig:gs_fiducial} -- though not quite as extreme -- have appeared in the literature before, most often a byproduct of restricting star formation to massive halos, assuming a dramatic suppression of X-ray emissions per unit star-formation in high-$z$ galaxies \citep[e.g.,][]{Pritchard2010a,Mesinger2013,Mirocha2015}, or assuming that the spectra of X-ray sources are very hard \citep{Fialkov2014} \footnote{Each of the aforementioned models, including ours, qualitatively agree when star formation is assumed to occur only in atomic cooling halos with a constant efficiency, independent of the modeling approach adopted. See, e.g., Figures 2, 9, 1, and 2 of \citet{Pritchard2010a}, \citet{Mesinger2013}, \citet{Mirocha2015}, and \citet{Fialkov2014}, respectively.}.

In contrast, the minimum mass of star-forming galaxies plays only a minor role in our framework \citep[as in][see also \S\ref{sec:Tmin_effects}]{Sun2016}. Our preference for models with strong high-frequency absorption features and weak (or non-existent) emission features is instead driven primarily by:
\begin{enumerate}
    \item The relatively inefficient $\fstar \sim 5$ \% peak star-formation efficiency and $\fstar \lesssim 1$\% in low-mass (but very abundant) halos, in contrast to the fiducial $\sim 10$\%.
    \item The steady reduction in $\tau_e$ over the last few years \citep{Planck2015,Planck2016}, which supports compressed reionization histories.
    \item Suggestive evidence that the high-$z$ X-ray background is dominated by X-ray binaries \citep{Lehmer2016,Brorby2016}, which have hard spectra and thus lead to relatively inefficient heating of the IGM.
\end{enumerate}
We are inclined to give our late-absorption (or ``cold reionization'') models more weight than was perhaps warranted in years past given their explicit calibration to the known high-$z$ galaxy population. \citet{Madau2016} recently came to a similar conclusion using up-to-date constraints on the star-formation history and a synthesis model for binary systems.

Given the recent interest in the spin temperature of the high-$z$ IGM, we show our predictions for the mean spin temperature at $z=8.4$ in the right-most inset of Figure \ref{fig:gs_fiducial}. The hatched region indicates unphysical values, while the grey region is disfavored by the \textit{PAPER} limits on the power spectrum \citep[the most pessimistic limits quoted in][]{Greig2016}. 

From an observational standpoint, the magnitude of changes in the global 21-cm signal brought about by changes in the SFE model are encouragingly large, as several forecasting studies have predicted that the absorption minimum of the global 21-cm signal can be recovered with an accuracy of $\lesssim 1$ MHz in frequency and $\lesssim 10-20$ mK in amplitude, at least for ideal instruments \citep[e.g.,][]{Pritchard2010a,Harker2012,Liu2013,Harker2016,Bernardi2016}. As a result, even a crude measurement of the global 21-cm signal could provide a much-needed constraint on the faint galaxy population in the early Universe. The goal of the remaining sections is to explore the degree to which uncertainties in other model parameters can complicate this procedure.

Before moving on, note that had we adopted the \citet{Finkelstein2015} LFs instead of the \citet{Bouwens2015} LFs our fiducial SFE would be systematically lower than that shown in Figure \ref{fig:calib_sfe} by about $\sim 15$\%. As a result, reionization and reheating would occur  later, and lead to a global 21-cm signal with even later features than those of Figure \ref{fig:gs_fiducial}. The impact of the calibration dataset can thus have important consequences, a point which we will revisit in Section \ref{sec:calib_improve}.

\subsection{Influence of the Minimum Mass of Star-Forming Halos} \label{sec:Tmin_effects} \label{sec:sfe}
Results shown in the previous section assumed that the minimum mass of star-forming galaxies is set by the atomic cooling threshold (return to Figure \ref{fig:calib_sfe} for a guide between $M_{\min}(z)$ and $\Tmin$). Of course, star-formation may proceed in lower mass halos via H$_2$ cooling, at least until a strong LW background develops \citep{Haiman1997}, or perhaps could be restricted to more massive halos if thermal feedback is effective \citep{Gnedin2000}. Though the mode of star-formation is not expected to be the same above and below the atomic threshold, we continue nonetheless in order to establish a reference case, devoting more thorough investigations including feedback to future work.

\begin{figure}
\begin{center}
\includegraphics[width=0.48\textwidth]{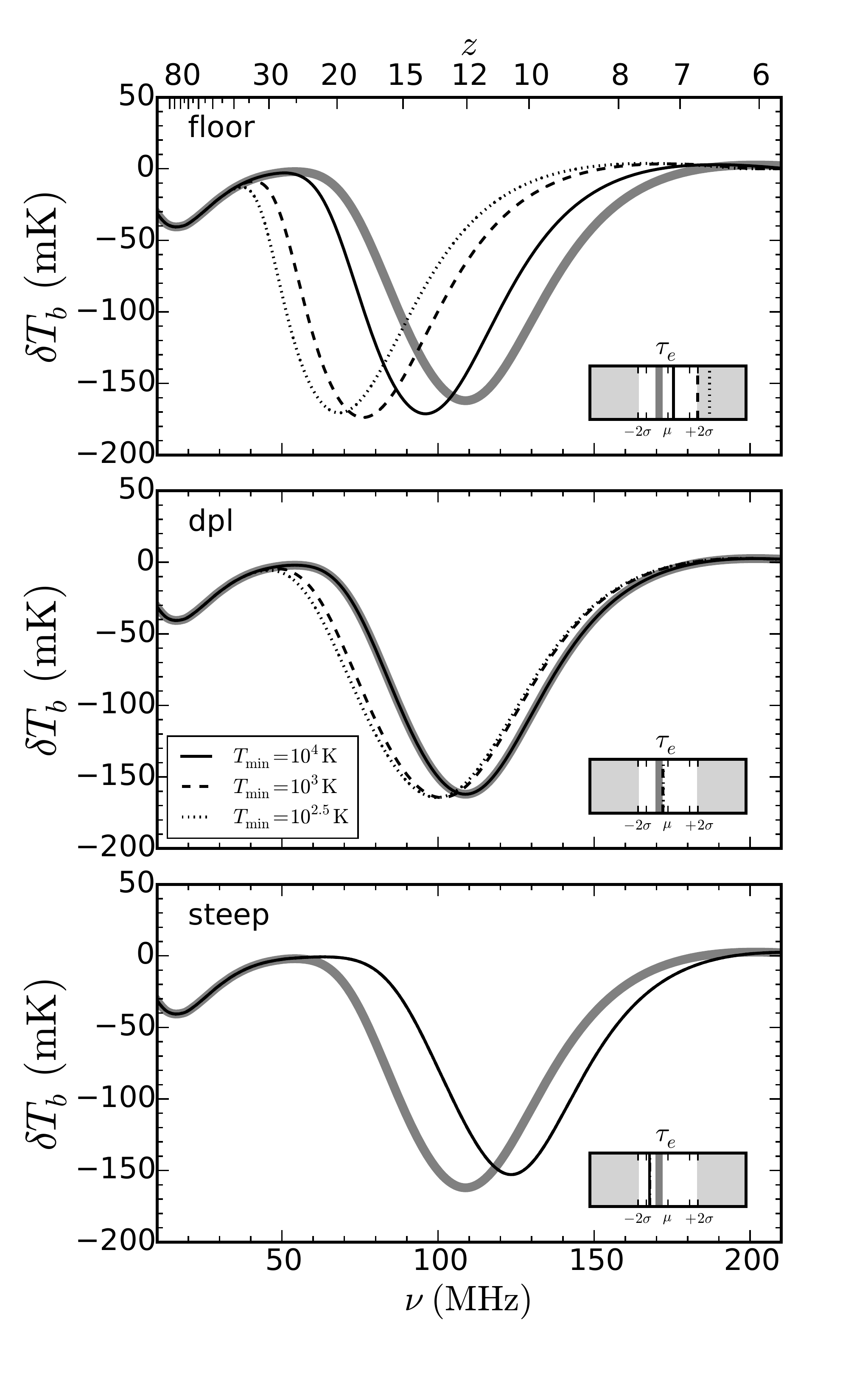}
\caption{Effects of the minimum virial temperature, for the \floor\ SFE model (top), double power-law model (middle), and \steep\ model (bottom). Different linewidths represent different $\Tmin$ choices, from the atomic cooling threshold (thickest lines) to molecular cooling halos (thinnest lines). Solid grey curve is our fiducial model and remains unchanged in each panel. Thomson scattering optical depth for all models is annotated in the lower right corner of each panel -- the light grey shaded regions are inconsistent with the recent \textit{Planck} measurements at the 95\% confidence level.}
\label{fig:Tmin_effects}
\end{center}
\end{figure}

In Figure \ref{fig:Tmin_effects}, we show how the global 21-cm signal depends on the low-mass SFE and $\Tmin$. The effects are most dramatic for the \floor\ SFE model (top panel), consistent with the \citet{Sun2016} result in the context of $\tau_e$ (their Figure 9). If $\Tmin = 10^4$ K, the signal shifts to lower frequencies by $\sim 10$ MHz and becomes deeper by $\sim 10$ mK in amplitude relative to our fiducial \dpl\ model. Decreasing $\Tmin$ to $10^3$ K, corresponding to halo masses of $\sim 10^6-10^7 \ \Msun$ shifts the signal by $\sim 20$ MHz toward lower frequencies, while adopting $\Tmin = 10^{2.5}$ K pushes the signal to lower frequencies by another $\sim 10$ MHz. These shifts are not without consequence, however, as the CMB optical depth for the $\Tmin \leq 10^3$ K models are inconsistent with the \textit{Planck} measurement at the  95\% confidence level (see inset). A reduction in the escape fraction or in the efficiency of stellar LyC photon production (relative to $1600$\AA\ production) would be required to resolve this discrepancy.

For our fiducial model (middle panel of Figure \ref{fig:Tmin_effects}), $\Tmin$ has a relatively minimal effect. At most, the absorption minimum shifts $\sim 10$ MHz relative to the fiducial case, with a negligible difference between the $10^{2.5}$ and $10^3$ K cases. The \dpl\ SFE is apparently steep enough that the contribution of very low-mass galaxies is negligible, effectively removing the importance of $\Tmin$. 

Lastly, invoking a \steep\ decline in the SFE (bottom panel of Figure \ref{fig:Tmin_effects}) shifts the global 21-cm signal absorption minimum to higher frequencies, as star-formation is restricted to the most massive -- and most rare -- halos. The same effect could be achieved by raising the virial temperature threshold to $\Tmin \gtrsim 10^5$ K, which could occur due to thermal feedback \citep{Gnedin2000}. However, note that in this case $\tau_e$ is approaching the $2-\sigma$ lower limit of \textit{Planck}. Boosting $\tau_e$ would require $\fescLyC \gtrsim 0.1$ or an enhancement in the LyC production efficiency (relative to $1600$\AA\ production). As a result, we will adopt $\Tmin = 10^4$ K in the sections that follow and investigate the impact of the stellar population model (\S\ref{sec:spop}) and the escape fraction (\S\ref{sec:escape}) quantitatively under this assumption.

\begin{figure*}
\begin{center}
\includegraphics[width=0.98\textwidth]{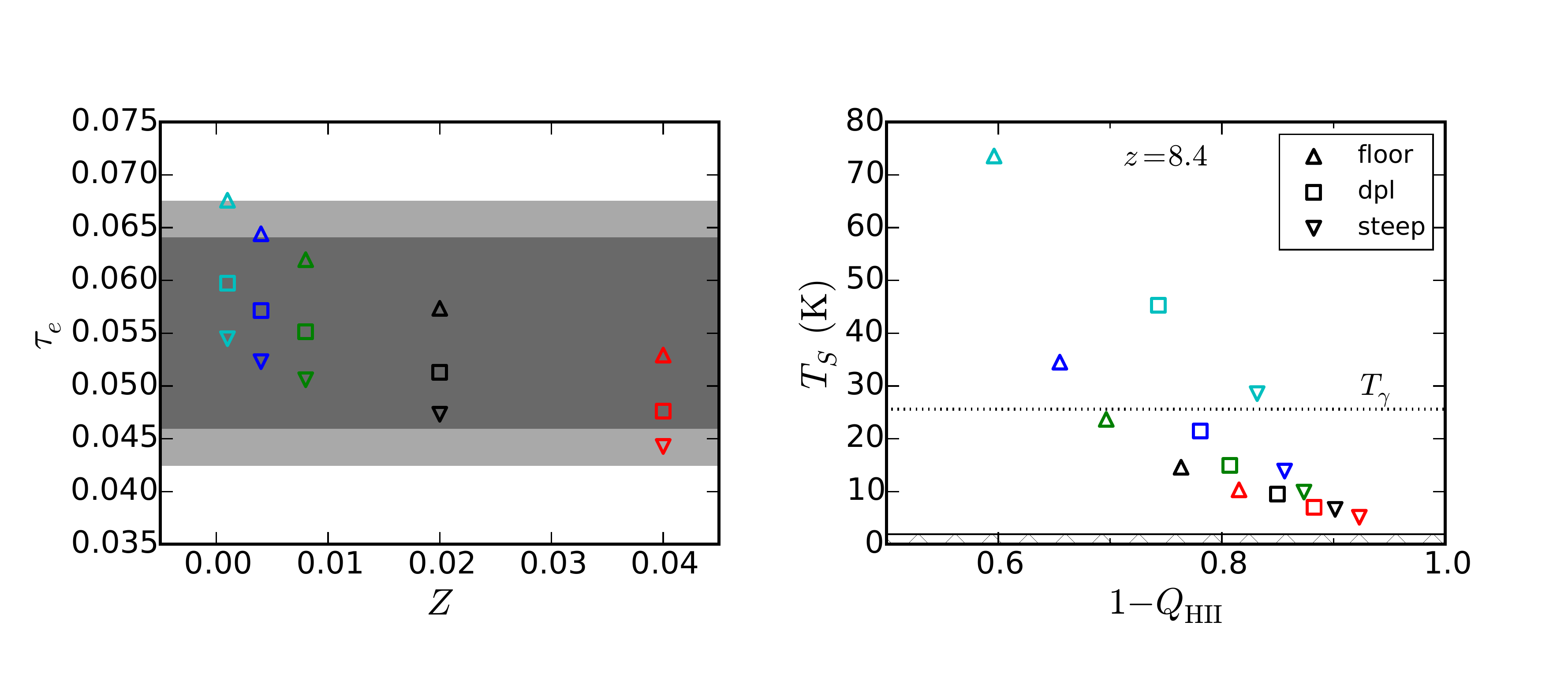}
\caption{Effects of the stellar metallicity and SFE model on the CMB optical depth and mean properties of the IGM at $z=8.4$. \textit{Left:} CMB optical depth as a function of stellar metallicity, relative to 68\% (dark grey) and 95\% (light grey) \textit{Planck} confidence regions. Different symbols represent the different SFE models, from the \floor\ SFE model (upper set of triangles), to \steep\ models (lower set of triangles). \textit{Right:} Mean IGM spin temperature vs. neutral fraction at $z=8.4$. Again, different colours correspond to different assumed stellar metallicities while the markers denote the different SFE models. The dotted line indicates the temperature of the CMB at $z=8.4$ and the narrow cross-hatched region at the bottom indicates unphysical temperatures. The \steep\ SFE models (downward triangles) with solar metallicity are disfavored by the recent \textit{PAPER} measurements \citep{Greig2016}.}
\label{fig:Ts_QHII_tau_Z}
\end{center}
\end{figure*}

\subsection{Stellar Population Effects} \label{sec:spop}
Figure \ref{fig:Tmin_effects} suggests that $\Tmin$ may still hold some power over the global 21-cm signal, though only if the SFE flattens (LF steepens) at low masses. Even in this case, $\Tmin$ primarily affects the \textit{timing} of the spectral features in the global 21-cm signal, with a much less dramatic influence on their amplitude. This may be encouraging -- perhaps then deviations from the amplitude of the trough in our reference model could provide clean evidence of other effects independent of its frequency. We focus in this section on how such deviations may reflect evolution in the stellar and black hole populations of high-$z$ galaxies, as parameterized through the metallicity of stellar populations, and the relationship between the metallicity and the X-ray luminosity star-formation rate ($L_X$-SFR) relation. 

Interestingly, and perhaps conveniently, \textit{the choice of stellar metallicity also impacts the inferred efficiency of star formation.} This is not because we have invoked any physical connection in our model, but because our SFE is calibrated to the galaxy LF, which requires an assumption about the $1600$\AA\ luminosity (per unit star formation). The result is that stellar metallicity has a smaller effect on the ionization history than one might naively expect.

For example, imagine we enhanced the $1600$\AA\ luminosity of galaxies (per unit SFR), $\mathcal{L}_{1600}$, by decreasing the stellar metallicity. Our model for the LF would shift systematically to the left, i.e., all galaxies would become brighter. To compensate, which we must do to preserve the best-fit to the observed LF, we would introduce a corresponding decrease in the SFE, which shifts the LF back toward the right (i.e., galaxies become fainter). Mathematically, we require:
\begin{equation}
    \fstar(Z_1) \mathcal{L}_{1600} (Z_1) = \fstar(Z_2) \mathcal{L}_{1600}(Z_2) . \label{eq:rhoL_conservation}
\end{equation}
While models with $Z_1$ and $Z_2$ have different star-formation histories, the redshift-evolution of the $1600$\AA\ luminosity \textit{density} is preserved\footnote{Assuming the functional form of the SFE and $\Tmin$ are independent of $\mathcal{L}_{1600}$.}. However, \textit{the luminosity density in other bands need not be preserved}. If, for example, the LyC luminosity of galaxies (per unit SFR), $\mathcal{L}_{\LyC}$, is more sensitive to $Z$ than $\mathcal{L}_{1600}$, then the ionization history will respond to $Z$ even if the LF remains fixed. Indeed, this is the case -- the \textit{shape} of the stellar spectrum changes with $Z$ -- meaning in general the ionizing background and LW background change with $Z$ even while holding the LF fixed. For example, the number of ionizing photons emitted per unit 1600\AA\ luminosity changes by a factor of $\sim1.7$ between the lowest and highest metallicities we consider (see Appendix \ref{appendix:spop}). If this were not the case, then LF-calibrated models for the global 21-cm signal would be completely insensitive to $Z$. 

In the left panel of Figure \ref{fig:Ts_QHII_tau_Z}, we show how metallicity and the SFE affect the CMB optical depth. It is immediately clear that each ($Z$, SFE) combination we consider leads to $\tau_e$ values consistent with \textit{Planck}. The spread in $\tau_e$ between our most extreme \floor\ and \steep\ SFE models (at fixed metallicity) corresponds roughly to the $1-\sigma$ \textit{Planck} uncertainty of $\sim 0.01$, which makes for a convenient rule of thumb. Similarly, for a given SFE model $\tau_e$ changes by $\sim 0.01$ between the lowest and highest metallicities we consider.

In the right panel of Figure \ref{fig:Ts_QHII_tau_Z} we show how metallicity and the SFE model affect the mean spin temperature and neutral fraction of the IGM at $z=8.4$, i.e., the same parameter space constrained recently by the 21-cm power spectrum limits from \textit{PAPER} \citep{Parsons2014,Ali2015,Pober2015}. Our \steep\ SFE models (downward triangles) are near the \textit{PAPER} limits (compare to Figure 1 of \citet{Greig2016}), though the rest are still well within the allowed region of parameter space. The majority of our models predict $T_S \leq \Tcmb$ at $z=8.4$, unless the SFE is pure double power law (as in our reference model) and the stellar metallicity takes on its minimal value, or there exists a floor in the SFE and $Z \lesssim 0.008$.

\begin{figure*}
\begin{center}
\includegraphics[width=0.98\textwidth]{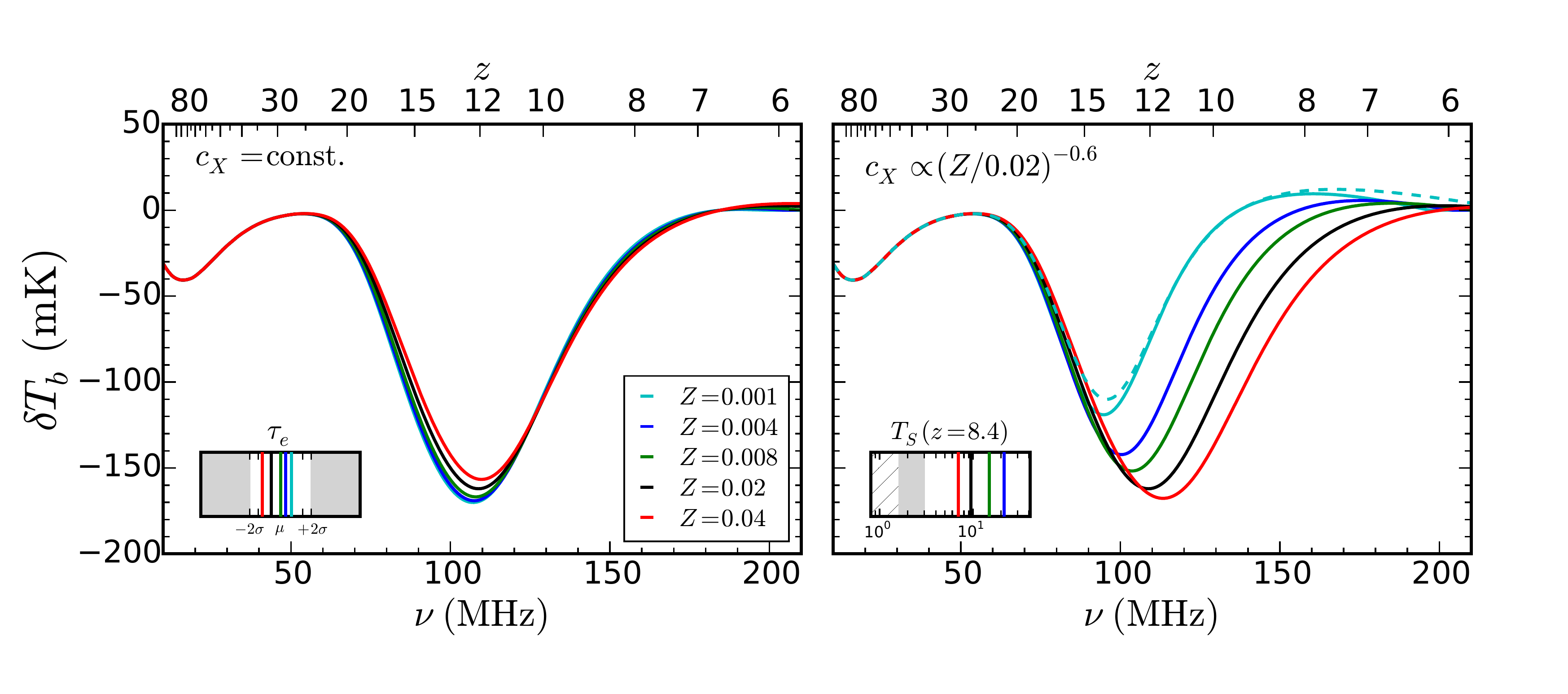}
\caption{Response of the global 21-cm signal to changes in the assumed stellar metallicity. \textit{Left:} Results obtained with a $Z$-independent $L_X$-SFR relation, which is equivalent to assuming the X-ray emission of galaxies is proportional to their MAR but not their stellar populations. \textit{Right:} Assumes $f_X \propto Z^{-0.6}$ $L_X$-SFR relation, as found in \citet{Brorby2016}. The dashed cyan curve shows the result obtained by assuming the $L_X$-SFR is boosted by a factor $(0.02 / Z)^{-0.6}$ but without actually changing $Z$, effectively decoupling $L_X$-SFR from the stellar metallicity. Again, insets show the corresponding CMB optical depth (left) and mean IGM spin temperature at $z=8.4$ (right).}
\label{fig:gs_Z} 
\end{center}
\end{figure*}

Changes in the spin temperature with metallicity can be quite large, up to a factor of $\sim 7$ in the most extreme \floor\ SFE model. This is partly a result of our assumption that the $L_X$-SFR relation depends on metallicity as $\mathcal{L}_X \propto Z^{-0.6}$ \citep{Brorby2016}, which we discuss in more detail below. The impact of $Z$ on $\QHII$ is much more modest, at $\sim 10-20$\%.

The efficiency of X-ray heating is often scaled by the parameter $f_X$, which modifies the locally-calibrated $L_X$-SFR relation \citep[e.g.,][]{Grimm2003,Gilfanov2004,Mineo2012a}. The heating rate density is then assumed to be a fixed fraction of the X-ray luminosity density, i.e.,
\begin{align}
    \eheat(z) & = f_X \left(\frac{\fheat}{0.2} \right) \left(\frac{2.6 \times 10^{39}}{\cXunits} \right) \nonumber \\  & \times \left(\frac{\rhostardot(z)}{\sfrdunits}\right) \coheatingdensity , \label{eq:LxSFR_heat}
\end{align}
where $\rhostardot$ is the star-formation rate density and $\fheat$ is the fraction of photo-electron energy deposited as heat \citep[as opposed to ionization or excitation;][]{Shull1985,Furlanetto2010}.

Our goal here is not to explore all possibilities for the $L_X$-SFR relation, which one can achieve by varying the normalization $f_X$ of Equation \ref{eq:LxSFR_heat} to arbitrary large or small values, but to explore changes consistent with our current understanding of star-forming galaxies. As a result, we use the metallicity-dependent $L_X$-SFR relation found by \citet{Brorby2016} as a guide. 

The expectation for some time has been $L_X \propto \mathrm{SFR} \times Z^{\beta}$, with $\beta< 0$, as low-metallicity environments ought to produce more massive stars and binaries \citep[e.g.,][]{Belczynski2008,Linden2010, Mapelli2010}. Observations of the Chandra Deep Field South \citep{BasuZych2013,Lehmer2016} find evidence of a boost in $L_X$-SFR with increasing redshift, which, interestingly, is close to the evolution allowed by the unresolved fraction of the cosmic X-ray background \citep{Dijkstra2012}. The \citet{Brorby2016} result implies that such evolution may simply reflect the metallicity evolution of galaxies, though measurements of the gas-phase metallicities in high-$z$ galaxies \citep[e.g.,][]{Sanders2016} will be needed to put this hypothesis to the test.

Rather than scaling $\eheat \propto f_X \rhostardot$ as in Equation \ref{eq:LxSFR_heat}, we solve the cosmological RTE in detail, which enables a more careful treatment of X-ray source SEDs \citep{Mirocha2014}. We use the MCD model both because it is representative of high-mass X-ray binaries, which are thought to be the most important sources in high-$z$ galaxies \citep[e.g.,][]{Fragos2013,Lehmer2016}, and also because their hard spectra provide a pessimistic limit in which heating is as inefficient as it could be\footnote{Note that inverse Compton emission from supernova remnants (SNRs) has an intrinsically harder spectrum than HMXBs \citep{Oh2001}, so our model could be even more pessimistic about the heating efficiency. However, over the $0.2 \lesssim  h\nu /\mathrm{keV} \lesssim 2$ keV interval in which most heating occurs, unabsorbed HMXB spectra and SNRs spectra are comparably flat.} (per 0.5-8 keV X-ray luminosity).

In the left panel of Figure \ref{fig:gs_Z} we first explore the effects of treating $f_X$ and $Z$ as completely independent (i.e., neglecting the empirically motivated variation of $f_X$ with $Z$ discussed above). Interestingly, the ordering of the curves runs counter to our typical intuition: decreasing $Z$ leads to \textit{less} efficient heating and thus deeper absorption troughs. This is because low-$Z$ stellar populations produce $1600$\AA\ photons more efficiently than metal-enriched populations, which requires star-formation to be less efficient if we are to match the luminosity function (see Equation \ref{eq:rhoL_conservation}). This leads to a systematic downward shift in the star-formation rate density, and thus the X-ray luminosity density and heating at all redshifts. This effect is also apparent in $\tau_e$ (see inset). 

Introducing a $Z$ dependent $f_X$ reverses this trend in the global signal\footnote{Note that while $\tau_e$ is sensitive to $Z$, it is insensitive to the $Z$-dependence of the $L_X$-SFR relation since HMXBs are a negligible source of ionization in the bulk IGM.}, since decreases in $Z$ and thus the SFE are compensated for by $f_X$. If $f_X \propto Z^{-0.6}$, as recently suggested by \citet{Brorby2016}, there is a spread of $\sim 50$ mK between the predicted absorption features. The dashed cyan curve shows the result one obtains by introducing a factor $(Z_{\min} / Z_{\max})^{-0.6} \simeq 9$ shift in $L_X$-SFR \textit{without} adjusting the normalization of the SFE (as in Equation \ref{eq:rhoL_conservation}). This may not seem like much: without the LF constraint, we would overestimate the heating and obtain an absorption trough only $\sim 10$ mK shallower than it should be. However, there is a much larger point to be made here: \textit{under our assumptions, metallicity can only account for a factor of $\sim 9$ change in the $L_X$-SFR relation}. As a result, detection of an absorption trough shallower than $\sim -110$ mK places strong constraints on either the metallicity dependence of the $L_X$-SFR relation or presence of additional sources of X-ray radiation in the early Universe\footnote{In general, the characteristic mass of BHs in X-ray binary systems may grow with decreasing $Z$, in which case the typical HMXB SED would become softer and result in more efficient heating. However, \citet{Mirocha2014} showed that a even factor of 100 change in $\Mbh$ results in only a $\lesssim 30$ mK change in the depth of the absorption minimum (their Figure 5), so treating $\Mbh=\Mbh(Z)$ is a secondary effect, at least when only stellar mass objects are imoprtant.}. We will explore one caveat to this result in the next section.

The actual metallicity of high-$z$ galaxies is not yet clear, though it does not seem unreasonable to assume sub-solar metallicity at $z > 6$ given the limited time available for chemical enrichment. Future constraints on $Z$ -- in addition to its dependence on galaxy properties like mass or star-formation rate -- could be readily incorporated into our model.

\subsection{Broadband Photon Escape} \label{sec:escape}
The fraction of LyC photons that escape from high-$z$ galaxies into the IGM has long been an important, but poorly constrained, quantity in reionization models. The linear dependence of the global 21-cm signal on the ionization history (see Equation \ref{eq:dTb}) suggests that a measurement of the global 21-cm signal can in principle constrain the escape fraction and UV production efficiency of high-$z$ galaxies. However, if reality is anything like our models, such inferences will prove difficult, as the signal has yet to saturate ($\TS \gg \Tcmb$) at late times in any of our models, which precludes a clean measurement of the ionization history without perfect knowledge of the spin temperature history.

\begin{figure*}
\begin{center}
\includegraphics[width=0.98\textwidth]{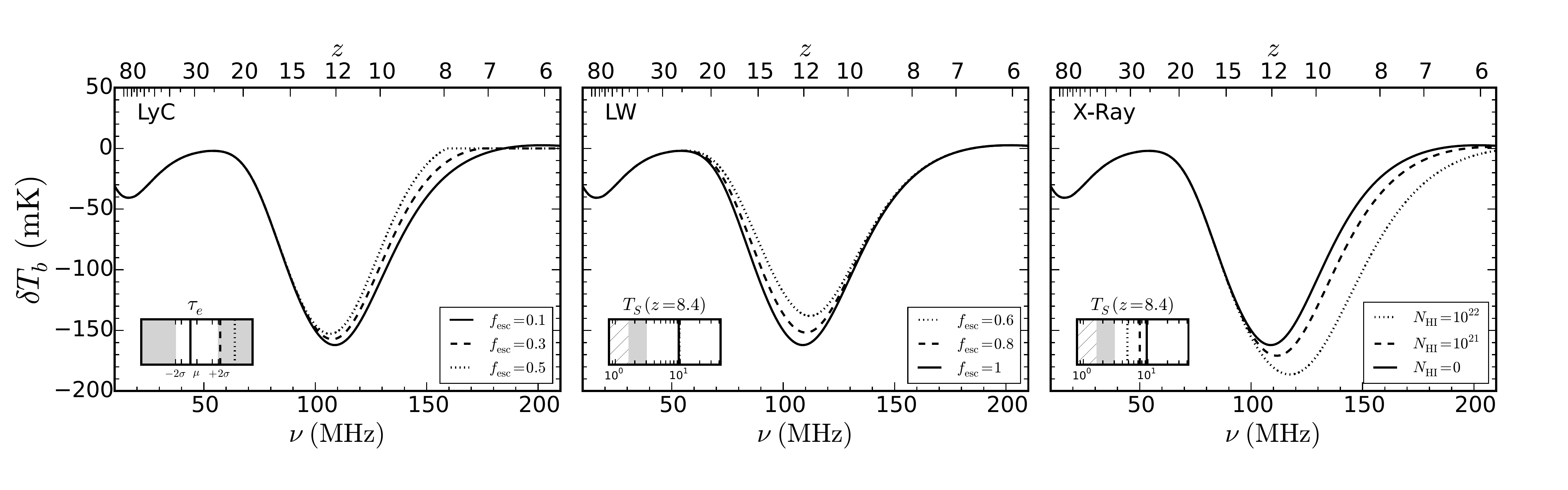}
\caption{Response of the global 21-cm signal to changes in the neutral gas contents of galaxies, including the fraction of LW photons that escape galaxies into the IGM (middle) the standard LyC escape fraction (left), and the characteristic neutral hydrogen column density in galaxies which hardens their X-ray emissions (right). Only the LyC escape fraction affects the CMB optical depth, which is shown in the left-most inset. Variations in the LW and X-ray escape affect the IGM spin temperature, which is shown in the insets of the middle and right panels.}
\label{fig:escape}
\end{center}
\end{figure*}

The left panel of Figure \ref{fig:escape} shows the global 21-cm signal's limited sensitivity to the LyC escape fraction. Increasing $\fescLyC$ from $\fescLyC=0.1$ to $\fescLyC=0.3$ pushes $\tau_e$ outside the \textit{Planck} 95\% confidence interval (see inset), but is barely discernible in the global 21-cm signal (main axes). This finding is not completely new: the high-frequency part of the global 21-cm signal is weak and fairly smooth spectrally \citep[see, e.g., Figure 1 of][]{Mirocha2015}, making it much more difficult to distinguish from the spectrally-smooth foregrounds than the strong expected absorption trough \citep{Harker2016}, even if the signal is fully saturated before reionization.

Though the prospects for constraining $\fescLyC$ are poor, the global 21-cm signal is still sensitive to the gas contents of high-$z$ galaxies. Absorption of photons by the Lyman series and Lyman-Werner bands could in principle be constrained by the global 21-cm signal, as it is photons in the $10.2 \leq h\nu \leq 13.6$ band which ultimately redshift or cascade through the $\Lya$ resonance and give rise to Wouthuysen-Field coupling between the spin temperature and kinetic temperature of the IGM.

As a result, a non-unity escape fraction of photons in this band would delay the onset of efficient coupling, and thus shift the absorption trough of the global 21-cm signal to higher frequencies and shallower depths, as shown in the middle panel of Figure \ref{fig:escape}. A 20\% change in $\fescLW$ from unity to $0.8$ results reduces the depth of the absorption trough by $\sim 10$ mK, with another 20\% decrease in $\fescLW$ leading to another $\sim 15$ mK shift in the absorption signal strength. These changes would be difficult to discern via other means as they affect neither the $z=8.4$ spin temperature (at a substantial level) or the ionization history (in contrast to comparable changes in $\fescLyC$ which push $\tau_e$ outside the \textit{Planck} confidence region)\footnote{Note that in general the LW background can have a more complex impact on the signal \citep[see, e.g.,][]{Fialkov2014b} given that it influences star-formation in minihalos before reionization \citep{Haiman1997}. However, here we focus only on atomic cooling halos in which LW feedback is unimportant.}.

To our knowledge, $\fescLW$ has been set to unity in every study of the 21-cm background to date. There is some theoretical support for $\fescLW < 1$, at least in idealized halos \citep[e.g.,][]{Kitayama2004,Schauer2015}, and observations that suggest a non-negligible H$_2$ opacity in star-forming galaxies \citep{Reddy2016}. Shallow $\dTb \lesssim -110$ mK absorption features could be achieved through variations in $\fescLW$, though would require $\fescLW \lesssim 0.4$, which is probably extreme \citep{Schauer2015}. To ``fill in'' the trough without invoking $\fescLW << 1$ we would need to introduce additional X-ray source populations or boost $L_X$-SFR with metallicity more strongly than suggested by \citet{Brorby2016}. Distinguishing $\fescLW < 1$ from efficient heating scenarios will require limits on the strength of the global 21-cm emission signal since, while both scenarios can cause weak absorption troughs, only efficient heating can drive a strong 21-cm emission feature. 

Lastly, X-ray sources are also subject to an escape fraction, parameterized by a neutral hydrogen absorbing column, $\NHI$ (see Equation \ref{eq:fescX}). Large absorbing columns harden the spectra of X-ray sources, which acts to lower the efficiency of X-ray heating and therefore produce stronger absorption troughs. We see in the right-most panel of Figure \ref{fig:escape} that a large $\NHI = 10^{22} / \mathrm{cm}^{2}$ absorbing column results in a $\sim -190$ mK trough and mean spin temperature of $\TS \sim 5$ K at $z=8.4$ (see inset), near the limit of what is allowed by \textit{PAPER}. In the limit of completely negligible heating, an absorption trough occurs at depths of $\lesssim -200$ mK due to the onset of reionization \citep[see, e.g., Figure 6 in][]{Mirocha2013}. 

\section{Discussion} \label{sec:discussion}
Our new model for the global 21-cm signal represents a conservative approach in which galaxies at arbitrarily high redshifts are assumed to share the same properties as those currently observed at $z \sim 6-8$, with some guidance from lower redshifts when necessary. Without PopIII stars, miniquasars, or other plausible-but-unconstrained objects, we predict a global 21-cm signal which peaks in absorption at $(\nu, \dTb) \sim (110 \ \mathrm{MHz}, -160 \ \mathrm{mK})$ and has virtually no emission signature at late times. 
Uncertainties in critical model parameters of course remain, but within the limits of current stellar population models, observational constraints on the reionization history, and assumption that the modeling formalism for the global 21-cm signal and galaxy LF themselves are appropriate, our main result is robust. This has important ramifications for observational efforts to detect and characterize the global 21-cm signal and theoretical efforts to interpret its meaning in the broader context of galaxy formation. 

Our goal in this section is to highlight the features and limitations of our model most pertinent to upcoming observations and future model development.

\subsection{Relevance to Global 21-cm Experiments} \label{sec:obs_consequences}

\subsubsection{Detection}
A persistent feature of our models is a deep absorption minimum at $\nu \sim 110$ MHz. If this prediction is accurate, it has important consequences for global 21-cm experiments in several regards. First, the FM radio band occupies $88-110$ MHz and will overwhelm even the strongest of signals. The FM band has typically not been a concern for experiments targeting the absorption minimum of the signal, since earlier predictions found a minimum closer to $\sim 70$ MHz and an emission maximum at frequencies of $\sim 115$ MHz, conveniently straddling one of the most important foreground contaminants. 

Even if the FM band can be avoided, e.g., by observing from the remote sites on the Earth's surface or the radio-quiet lunar farside \citep{Burns2012}, separation of the signal from the foreground could be more difficult if it lies near the edge of one's band of observation. If, for example, only half of the broad absorption feature falls within band its spectral complexity is reduced, and the signal could thus be more easily mimicked by the galactic foreground, especially given the complex effects of realistic beams \citep{Bernardi2015}. As a result, a broad-band is optimal, as advocated by \citet{Mirocha2015} (albeit for reasons of interpretation rather than detectability), though simultaneously fitting data taken in disjoint frequency intervals may be an economical method for obtaining broad-band constraints.

While extracting the full global 21-cm signal is the ultimate goal for experiments, model rejection techniques can be readily applied today. To date, the only constraints on the emission feature come from \citet{Bowman2010}, who, operating under the assumption of a saturated signal, were able to rule out the sharpest reionization scenarios that would manifest as a $\sim 20$ mK ``step'' relative to the smooth galactic foreground. Our findings suggest that an analogous exercise invoking a deep trough rather than a sharp emission step could be fruitful. 

Indeed, \citet{Bernardi2016} recently performed such a test with $\sim 20$ minutes of $50 \leq \nu / \mathrm{MHz} \leq 100$ LEDA data, limiting the depth of the absorption minimum to $\delta T_{b,\min} \gtrsim -1$ K, which lies within a factor of $\sim 3$ of the strongest absorption signal that is physically allowed. In the near future, such an approach may be used to rule out some of our models. Luckily, the strongest signals considered in this study are also the most ordinary, having solar metallicity, a \dpl\ SFE, and unattenuated LW and X-ray emissions, meaning evidence for non-standard prescriptions might emerge sooner rather than later. 

\subsubsection{Interpretation} \label{sec:selection_criteria}
Throughout, we have emphasized that our model is as conservative as possible in that we only include known sources. By identifying the realizations of the global 21-cm signal consistent with the status quo, set today primarily by galaxy LF measurements and $\tau_e$, we effectively define a null test for a global 21-cm signal detection. That is, if \textit{all} of the following statements are in the future observed to be true, then no dramatic changes to the model are required:
\begin{enumerate}
    \item The absorption trough occurs at frequencies $\nu \gtrsim 90$ MHz.
    \item The absorption trough is stronger than $\nu \sim -110$ mK.
    \item The emission maximum is weaker than $\sim 10$ mK.
\end{enumerate}    
If any of the above statements are false, our model is either wrong or incomplete. 

Realizations of the global 21-cm signal inconsistent with \#1 above could occur if the SFE flattens at low masses (LF steepens at faint end; see Figure \ref{fig:Tmin_effects}) or if PopIII star formation in minihalos (which we have neglected) is efficient. Both scenarios would trigger Wouthuysen-Field coupling at earlier times than our other models predict, but simultaneously enhance the ionizing background. As a result, a reduction in $\fesc$ would likely be required to maintain consistency with the \textit{Planck} constraint on $\tau_e$. 

Very shallow troughs (contrary to item \#2 above) can be achieved by very small LW escape fractions or additional heat sources \citep[e.g., PopIII remnants or direct collapse BHs;][]{Tanaka2016}. The argument for additional heat sources, rather than $\fescLW < 1$, would become stronger if the signal had a shallow trough \textit{and} a strong emission signal, violating items 2 and 3 above. 

Finally, all three criteria are likely broken if Population III star formation is efficient and their remnants accrete persistently. However, the details are complex, as PopIII star formation is subject to potentially several feedback mechanisms, such as H$_2$ destruction (or catalysis) induced by large-scale radiation backgrounds \citep[e.g.,]{Haiman1997,Ricotti2016}, and the relative supersonic motion between baryons and dark matter after recombination \citep{Tseliakhovich2010}. Given that our goal is to establish a conservative reference case, we leave a more thorough investigation of PopIII scenarios to future work.

\subsection{Prospects for Joint LF -- Global 21-cm Inference} \label{sec:LFGS}
The criteria outlined in \ref{sec:selection_criteria} suggest that even a relatively crude initial detection should be able to distinguish a global 21-cm signal that is broadly consistent with our model from one that is not. However, the differences between models in our study are also large enough that -- should similar realizations be observed -- precision measurements could be used to constrain interesting parameters like the stellar metallicity, $L_X$-SFR relation, and SFE in low-mass galaxies, which is equivalent to the faint-end slope of the galaxy LF in our formalism.

Given that the location of the global 21-cm absorption minimum varies by $\sim 30$ MHz between the extreme \floor\ and \steep\ SFE models we consider, the prospects for using the global 21-cm signal to probe faint galaxies are encouraging. Comparable shifts in the minimum can, however, arise from other effects, which is of some concern (for example lowering the halo mass threshold for star formation (see \S\ref{sec:Tmin_effects}) or invoking a redshift-dependent SFE (see \S\ref{sec:calib_improve})). This means commensurate progress in LF and 21-cm measurements will likely be required to distinguish changes in the SFE at low-mass from $\Tmin < 10^4$ K scenarios. In contrast, the metallicity and $L_X$-SFR relation (\S\ref{sec:spop}) and LW and X-ray escape fractions (\S\ref{sec:escape}) primarily affect the amplitude of the trough and emission signal, and thus may be constrained independently of the details of the SFE and $\Tmin$.

\subsection{Relevance to Power Spectrum Experiments} \label{sec:PS}
Though we have not attempted to model the 21-cm power spectrum explicitly, our results still have implications for power spectrum experiments. 

Our fiducial model predicts a mean IGM spin temperature of $\TS \sim 10$ K at $z = 8.4$ and neutral fraction of $1-\QHII \sim 0.85$. For comparison, limits for these quantities were recently published by \citet{Pober2015} and \citet{Greig2016} in response to improving upper limits on the power spectrum from \textit{PAPER} \citep{Parsons2014,Ali2015}. \citet{Greig2016} find that $\TS \gtrsim 3$ K assuming a neutral fraction $>10$\%, $\TS \gtrsim 5$ K if one tightens the assumed neutral fraction range to 30-65\%, and to $\TS \gtrsim 6$ K if one includes priors on the IGM ionization state at $z = 5.9$ \citep{McGreer2015} and $\tau_e$ \citep{Planck2015}. If the IGM is cold during reionization, the assumption of saturation -- typically used to boost the computational efficiency of power spectrum calculations -- will be poor, and may require such constraints to be revisited in more detail.
 
In this study, we have not attempted to incorporate the \textit{PAPER} constraints directly in our fitting since we do not generate a new model for the ionization and spin temperature histories on each MCMC step. Instead, we only use MCMC to fit the galaxy LF, from which point we take the best-fit SFE parameters and run global 21-cm signal models separately. However, it seems clear that simultaneously considering multiple measurements at once could be very powerful. For example, \citet{Liu2016b} recently showed that knowledge of the power spectrum can aid detection of the global signal. On the other hand, an independent measurement of the global 21-cm signal would identify the source of the fluctuations underlying the power spectrum.

Though forecasted constraints on the model parameters obtained by power spectrum measurements were found to only improve modestly upon inclusion of the global signal, \citet{Liu2016b} point out that this is likely due to the weakness and spectral smoothness in the global 21-cm emission feature, which was the only piece of the signal employed in their study. The 21-cm absorption signal is likely to be much more powerful in such a role given the large shifts (tens of MHz in frequency and tens of mK in amplitude) that result from reasonable changes in model parameters of interest. 

\subsection{Improving the Model Calibration} \label{sec:calib_improve}
Current uncertainties in LF measurements have a noticeable impact on our results, as evidenced by the spread in $\tau_e$, $\TS(z=8.4)$, and global 21-cm signal induced by our different SFE models (\floor, \dpl, \steep), which are meant to roughly span the range of possibilities in the (currently unconstrained) faint-end of the LF. There are two ways of looking at this: (i) the allowed range of models means global 21-cm signal measurements can help constrain the galaxy LF, and (ii) improved LF measurements will help to further refine our range of ``vanilla'' global signal models. We will explore the latter view more thoroughly below, as we have already addressed point (i) in \S\ref{sec:LFGS}.

\begin{figure*}
\begin{center}
\includegraphics[width=0.98\textwidth]{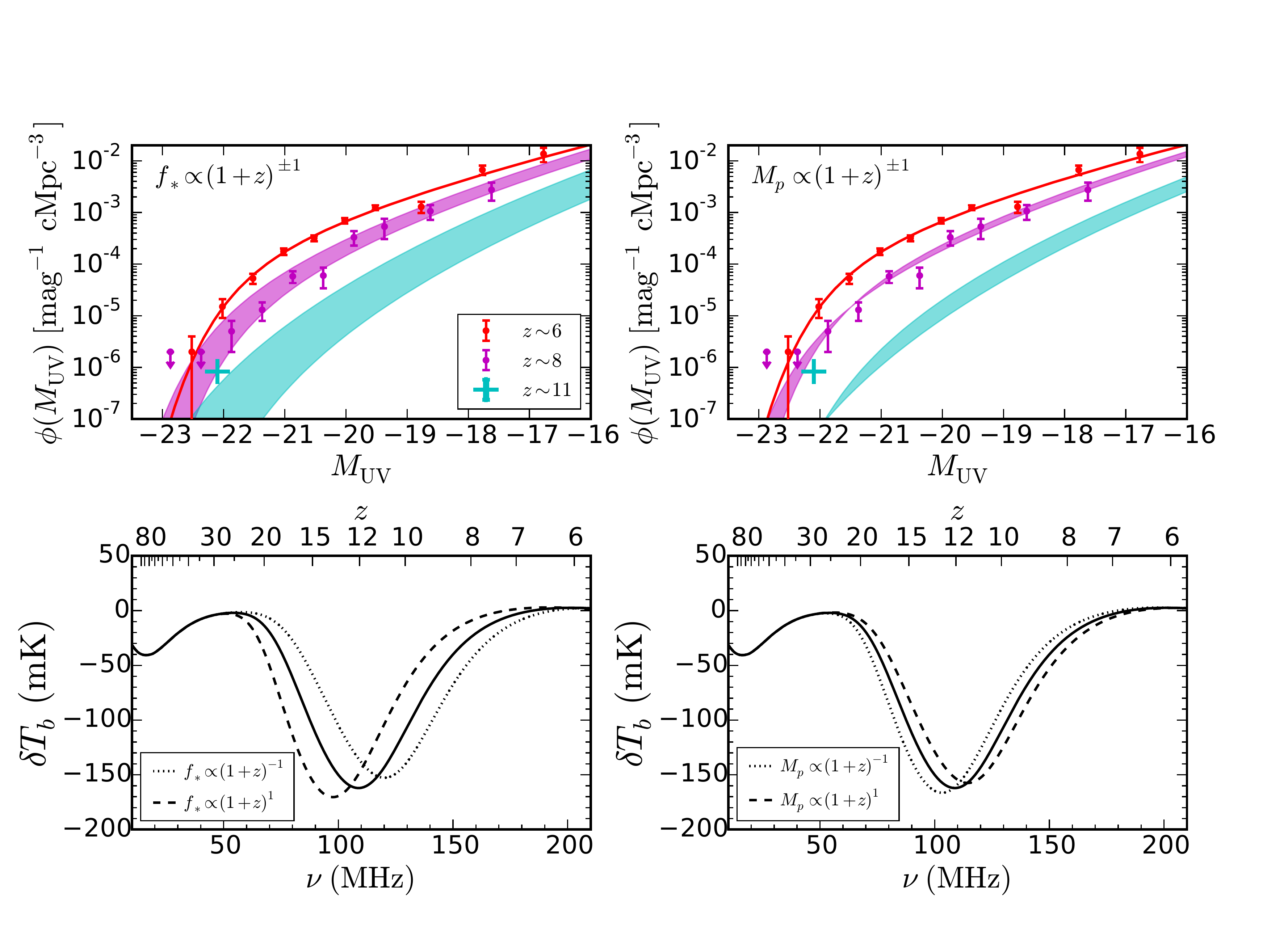}
\caption{Effects of redshift evolution in the SFE on the high-$z$ galaxy LF and global 21-cm signal. \textit{Left:} Redshift evolution in $f_{\ast}(M_p)$ is apparent in the LF (top) and global 21-cm signal (bottom). \textit{Right:} Results obtained assuming the peak mass, $M_p$, rather than the peak normalization, $f_{\ast}(M_p)$, is varied as $(1+z)^{\pm 1}$. In the upper panels, $z \sim 6$ and $z \sim 8$ data points are from \citet{Bouwens2015}, while the $z \sim 11$ point is the object discovered by \citet{Oesch2016}.}
\label{fig:lfgs_zdep_sfe}
\end{center}
\end{figure*}

Extending LF constraints to fainter magnitudes would clearly influence our model's calibration, as is likely to occur in the near-term via \textit{JWST}. Systematic errors are also a concern for our calibrated models, as for example, the $z \sim 6-7$ \citet{Bouwens2015} and \citet{Finkelstein2015} LFs are discrepant by a factor of $\sim 2$ at $\MUV \lesssim -18$, which is larger than the quoted statistical uncertainties in the measurements. Hopefully, future measurements will alleviate such tensions. For now, we note that adopting the \citet{Finkelstein2015} LF, rather than that of \citet{Bouwens2015}, would only strengthen our conclusions, since the \citet{Finkelstein2015} LFs imply even less efficient star-formation in high-$z$ galaxies, meaning it is even more difficult to make shallow absorption troughs and/or strong emission features in the global 21-cm signal.

Additionally, we have not attempted to use any $z < 6$ galaxy LF data. Doing so may provide evidence for redshift evolution in $\fstar$, but such inferences are complicated by the increasing importance of dust at $z \lesssim 6$ \citep{Bouwens2012,Capak2015}. For now, we present a few plausible extensions to the SFE in Figure \ref{fig:lfgs_zdep_sfe} in which its normalization and peak mass evolve with redshift.

First, we simply scale the normalization of the SFE as a power-law in redshift (left panels of Figure \ref{fig:lfgs_zdep_sfe}),
\begin{equation}
    f_{\ast}(M_p) = f_{\ast,0} \left(\frac{1 + z}{7}\right)^{\gamma_{\ast}}
\end{equation}
and allow the power-law index to vary between $-1 \leq \gamma_{\ast} \leq 1$. This causes a $\sim 0.2-0.5$ dex change in $\phi(\MUV)$ at $z \sim 8$ (magenta band; top left panel), growing to $\sim 0.5-1$ dex at $z \sim 11$ (cyan band). The absorption minimum of the global signal varies by $\sim \pm 10$ MHz in position, and in amplitude by $\sim 10$ mK.

Next, we fix the amplitude of the SFE curve but allow the location of the peak mass to vary with redshift as
\begin{equation}
    M_p = M_0 \left(\frac{1 + z}{7}\right)^{\gamma_M}
\end{equation}
with $-1 \leq \gamma_M \leq 1$, and $M_0 = 3 \times 10^{11} \Msun$ as in our fiducial model. The results are shown in the right panels of Figure \ref{fig:lfgs_zdep_sfe}, and exhibit a smaller spread in the LF (top right) and global signal (bottom right) than did evolution in $f_{\ast}(M_p)$. The LF changes very little between $-1 \leq \gamma_M \leq 1$, while the minimum of the global 21-cm signal varies by $\pm 5$ MHz, in contrast to the changes of order $\pm 10$ MHz caused by the $f_{\ast}(M_p)$ evolution (lower left panel). This model struggles to produce $\MUV \sim -22$ galaxies with an abundance similar to that implied by the $z \sim 11.1$ galaxy recently discovered by \citep{Oesch2016}, but given that this point represents just one object we caution the reader against over interpretation of these findings.

In addition to using longer redshift baseline in LF measurements, future studies could leverage stellar mass functions \citep[e.g.,][]{Song2016} or constraints on the UV luminosity density at $z \sim 4-5$ from the $\Lya$ forest \citep[e.g.,][]{Becker2013}. 

Figure \ref{fig:Ts_QHII_tau_Z} shows that reducing uncertainties in $\tau_e$ could also have an important impact on our calibration, at least in principle, as the $1-\sigma$ error on $\tau_e$ roughly corresponds to the differences in our model brought about by the uncertain behavior of the SFE at low mass. However, without commensurate progress in our understanding of UV photon escape from galaxies, any updates to $\tau_e$ can be ascribed solely to $\fescLyC$, rather than to the SFE, to which the global 21-cm signal is more sensitive. This may actually be a blessing in disguise: it implies that the global 21-cm signal can be used to infer galaxy properties at high-$z$ without understanding $\fescLyC$ at all. This is in stark contrast to efforts to reconcile $\tau_e$ and the galaxy LF, in which case $\fescLyC$ plays a central role.

The global 21-cm signal and/or power spectrum themselves can in principle provide an independent measurement of $\tau_e$ \citep{Pritchard2010b,Liu2016a,Fialkov2016a}, though if reality is anything like our fiducial model, this will prove to be extremely difficult as the extraction of $\tau_e$ relies on a strong emission signal (at least for the global 21-cm signal), which we predict to be weak or nonexistent. This serves to emphasize the intimate link between the ionization and thermal histories in the 21-cm background, and how constraints on one are only as good as constraints on the other.

Finally, improved constraints on the X-ray emissions of high-$z$ galaxies could be readily incorporated into our models, e.g., measurements of the $L_X$-SFR relation, including its redshift evolution, metallicity dependence, and potentially additional scalings (e.g., with the stellar mass or SFR itself). Further examination of the types of X-ray sources inhabiting high-$z$ galaxies would either help substantiate our decision to only include X-ray emissions from HMXBs, or force us to abandon it. The unresolved fraction of the cosmic X-ray background might provide some additional help in constraining high-$z$ X-ray  emissions, but based on the results of \citet{Fialkov2016b}, the unresolved XRB will only provide a useful constraint if $f_X$ is very large.

\subsection{Extending the Model}
Because our models are more restrictive than, e.g., the four parameter model of \citet{Mirocha2015} or the tanh model of \citet{Harker2016}, their use in signal extraction as-is may not be warranted. For example, in the event that our model is dissimilar from the true signal, its rigidity would prevent a good fit and instead cause large biases to result in the posterior distributions of model parameters (unbeknownst to us). It would be preferable to either (i) first detect and characterize the signal using a flexible and efficient model, and follow-up by comparing to more detailed (but expensive) models like ours, or (ii) augment our model with new source populations, to be described below.

Additions to the model might include new sources of $\Lya$, LyC, and X-ray photons such as PopIII stars, miniquasars, or even more exotic candidates such as annihilating or decaying DM. By fitting for, e.g., the formation efficiency of PopIII stars, it would be possible to quantify the need for PopIII stars by the data. This has the appearance of a high-level model selection exercise to follow initial null tests, such as those proposed in \citet{Harker2015}, in which our fiducial LF-calibrated model would provide a common basis from which to test the necessity of new extensions. 

Aside from augmenting the model with entirely new components, further insights regarding the relationship between its underlying components could also be interesting. For example, we have not made any attempt to link $\fescLW$, $\fescLyC$, and $\NHI$ in a physical model, nor have we invoked $M_h$-dependent escape fractions or stellar populations. Further study of such effects seem warranted given the potential impact on the global 21-cm signal and galaxy LF. Guidance from numerical simulations would be most welcome in these areas, as detailed modeling is beyond the scope of this work.

\section{Conclusions} \label{sec:conclusions}
We have shown that linking models of the global 21-cm signal to recent measurements of the high-$z$ galaxy LF leads to a strong preference for models with late heating and reionization ($z \lesssim 12$). At the formalism level, we assume that the $\fstar$-based model is the true model for the galaxy luminosity function, that the stellar population synthesis models we employ are accurate, and that the IGM is reasonably well-modeled as a two-phase medium. As for the components of the model, we assume that high-mass X-ray binaries are the sole sources of the $z \gtrsim 6$ X-ray background, and follow a $L_X$-SFR relation similar to that of local star-forming galaxies. If these assumptions hold, then our model suggests that:
\begin{enumerate}
    \item The global 21-cm signal peaks in absorption at $\nu \sim 110$ MHz and a depth of $\sim -160$ mK. The emission feature is negligible in most models, reaching an amplitude $\lesssim 10$ mK at $\nu \sim 150$ MHz in the most optimistic scenario (Figures \ref{fig:gs_fiducial} and \ref{fig:gs_Z}). Ruling out such models may be easier than the sharp step-function emission models typically targeted at $\nu \gtrsim 100$ MHz, and would provide clear evidence of non-standard physics and/or source populations.
    \item The absorption trough in the global 21-cm signal is very sensitive to the SFE in low-mass galaxies (Figure \ref{fig:gs_fiducial}), with a $\sim 30$ MHz spread in its position arising from differences between currently viable models. Constraining the SFE will be very difficult with the LF alone given the depths one must probe ($\MUV \sim -12$) to differentiate models.
    \item The minimum mass of halos capable of supporting star formation, parameterized through $\Tmin$, has only a minor impact on our results (Figure \ref{fig:Tmin_effects}) given the steep decline of the SFE with halo mass implied by the LF (Figure \ref{fig:calib_sfe}). 
    \item The $Z$-dependence of the $L_X$-SFR relation is very important, affecting the depth of the absorption feature at the $\sim 50$ mK level (Figure \ref{fig:gs_Z}), which corresponds to mean IGM spin temperatures between $\sim 10$ and $40$ K at $z=8.4$ (Figure \ref{fig:Ts_QHII_tau_Z}), close to the recent \textit{PAPER} constraints. Stellar metallicity plays a relatively minor role in setting the ionization history (Figure \ref{fig:Ts_QHII_tau_Z}). 
    \item The global 21-cm signal is not very sensitive to the LyC emission properties of galaxies. However, the escape of LW and X-ray photons may have a dramatic impact on the signal (Figure \ref{fig:escape}).
\end{enumerate}

J.M. would like to thank Louis Abramson for many stimulating conversations and comments on an earlier draft, and the anonymous referee for comments that helped improve this paper. This work was supported by the National Science Foundation through award AST-1440343 and by NASA through award NNX15AK80G. We also acknowledge a NASA contract supporting the ``WFIRST Extragalactic Potential Observations (EXPO) Science Investigation Team'' (15-WFIRST15-0004), administered by GSFC. SRF was partially supported by a Simons Fellowship in Theoretical Physics and thanks the Observatories of the Carnegie Institute of Washington for hospitality while much of this work was completed. 

\bibliography{references}
\bibliographystyle{mn2e_short}


\appendix

\section{Stellar Population Models} \label{appendix:spop}
Given the impact of stellar metallicity variations on the thermal and ionization histories, and thus global 21-cm signal (see Figures \ref{fig:Ts_QHII_tau_Z} and \ref{fig:gs_Z}), we thought it appropriate to test how the stellar population synthesis model itself affected our main results. 

Throughout the main text we used the \textsc{BPASS} version 1.0 models  \citep{Eldridge2009} \textit{without} binary evolution to derive the UV luminosity (per unit star formation) in the Lyman continuum band, the $10.2 \leq h\nu /\mathrm{eV} \leq 13.6$ band (loosely referred to as the LW band), and the $1600$\AA\ monochromatic intensity. We computed these values from assuming continuous star formation in the steady-state limit when $t \gtrsim 100$ Myr and neglect nebular emission.
 
\begin{figure}
\begin{center}
\includegraphics[width=0.48\textwidth]{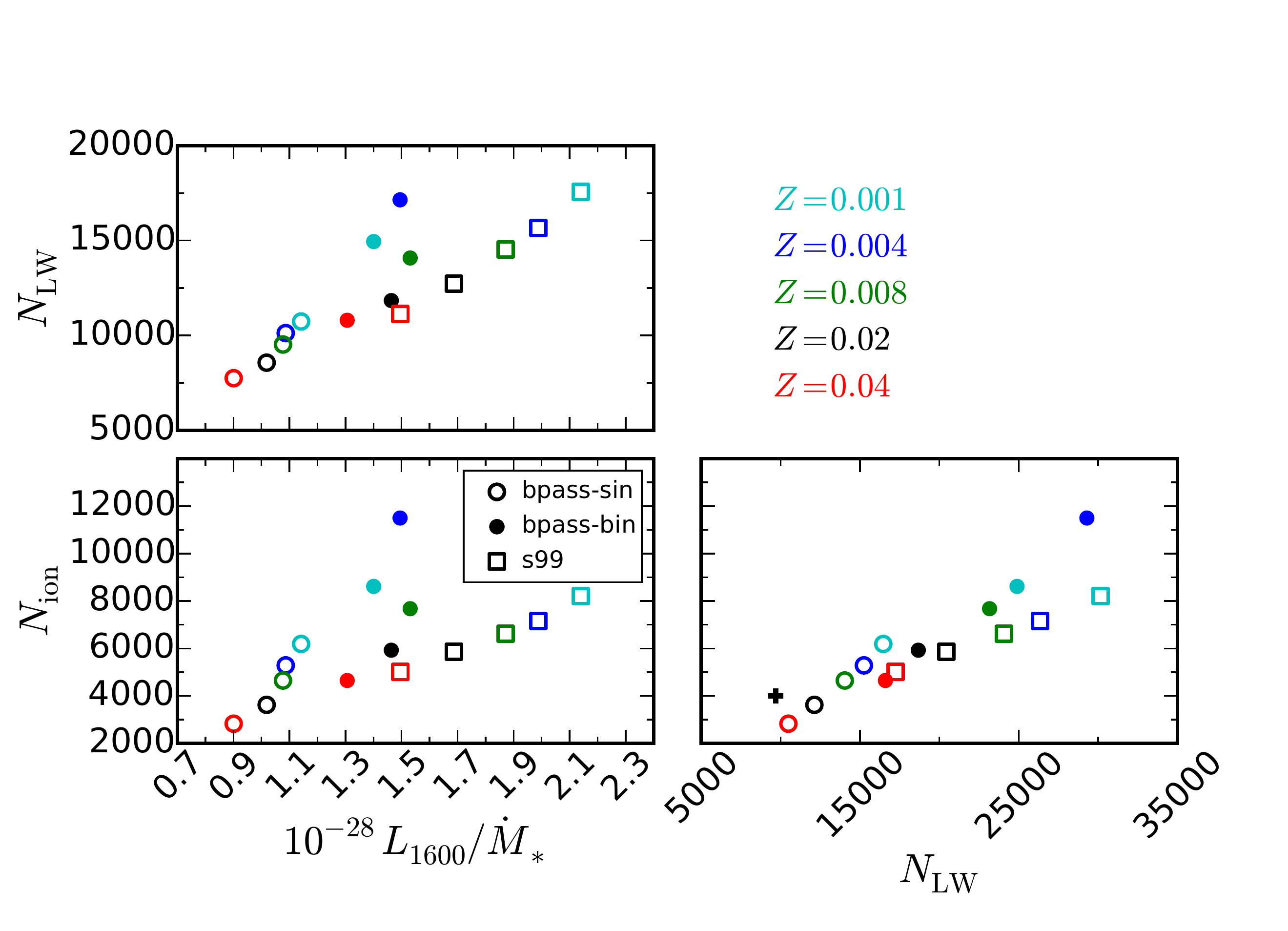}
\caption{Relationship between $\Nion$, $\Nlw$, and $\mathcal{L}_{1600}$ colour-coded by stellar metallicity for single star models from \textit{BPASS} (open circles) and binary models from \textsc{BPASS} (filled circles), and \textsc{starburst99} (squares). The cross in the lower-right panel denotes commonly adopted values from the literature based on earlier stellar population calculations with \textsc{starburst99} assuming a Scalo IMF \citep{Barkana2005}.}
\label{fig:spop}
\end{center}
\end{figure}
 
Figure \ref{fig:spop} shows the relationship between the LyC yield ($\Nion$; integrated from 1-2 Ryd), $10.2 \leq h\nu /\mathrm{eV} \leq 13.6$ yield ($\Nlw$), and $1600$\AA\ intensity for our fiducial models, as well as the \textsc{BPASS} version 1.0 models \textit{with} binaries and the original \textsc{starburst99} models \citep{Leitherer1999}. Yields in each band are strongly correlated with each other, which is to be expected given their close association in photon wavelength. The \textsc{starburst99} models (squares) predict somewhat larger values of $\mathcal{L}_{1600}$ than does \textsc{BPASS}, but the the spread of all models is a factor of $\sim 2-3$.

\begin{figure*}
\begin{center}
\includegraphics[width=0.98\textwidth]{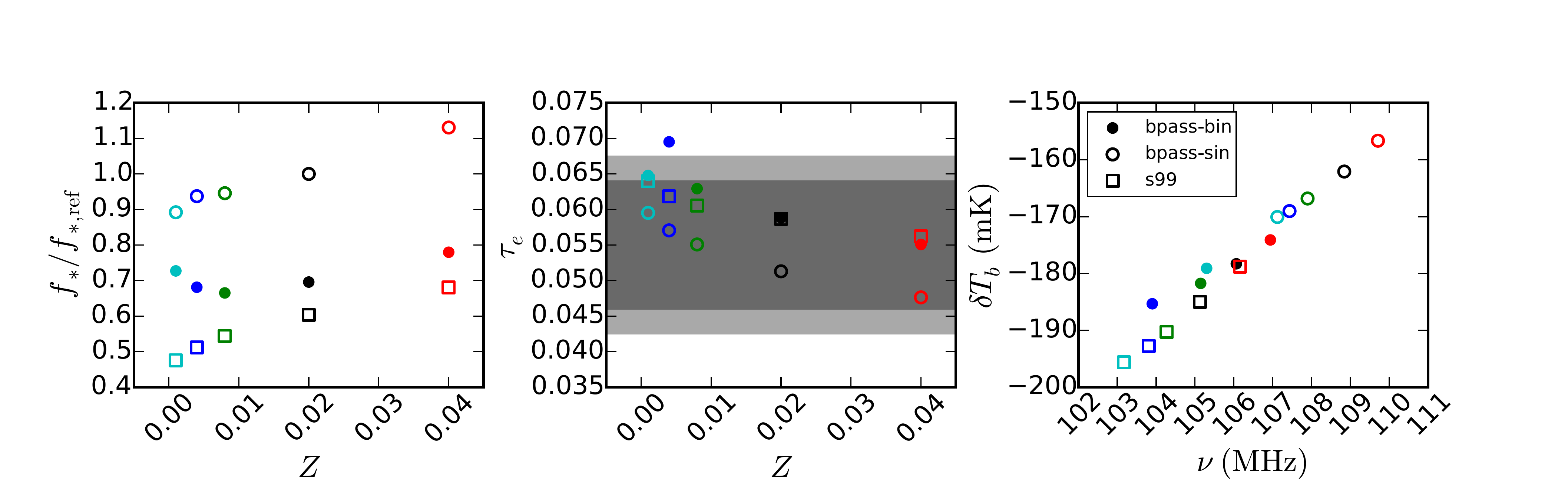}
\caption{Effects of the adopted stellar population synthesis model. \textit{Left}: Variation in the peak SFE, which results from each stellar population models differing values for $\mathcal{L}_{1600} / \dot{M}_{\ast}$. \textit{Middle}: CMB optical depth as a function of metallicity and stellar population model. \textit{Right}: Variation in the position of the global 21-cm absorption trough as a function of metallicity and stellar population model. }
\label{fig:spop_effects}
\end{center}
\end{figure*}

Figure \ref{fig:spop_effects} shows how these variations in the stellar population affect the normalization of the SFE, the CMB optical depth, and the location of the absorption minimum of the global 21-cm signal. Equation \ref{eq:rhoL_conservation} guarantees a correlation between the normalization of $\fstar$ and $Z$, which spans the same factor of $\sim 2-3$ as seen in $\mathcal{L}_{1600}$ in Figure \ref{fig:spop_effects} by definition. There is \textit{not} a corresponding factor of $\sim 2-3$ change in $\tau_e$, however. This is because $\fstar$ and $\mathcal{L}_{1600}$ are anti-correlated, as described previously. Due to the correlation between $\Nion$ and $\mathcal{L}_{1600}$, $\Nion$ and $\fstar$ are anti-correlated as well, and so for a given stellar population model, the dependence of $\tau_e$ on $Z$ is fairly weak. The spread between the models is comparable to the size of the $1-\sigma$ \textit{Planck} error bar.

The choice of stellar population model also affects the global 21-cm signal. In the right panel of Figure \ref{fig:spop_effects}, we show how the position of the absorption minimum changes as a function of the SFE model and stellar metallicity. All models occupy a relatively narrow range in frequency, $103 \lesssim \nu_{\min} / \mathrm{MHz} \lesssim 110$, though they span $\sim 50$ mK in amplitude. For each stellar population model, the spread due to metallicity is much smaller, at $\sim 3-4$ MHz in frequency and $\sim 10-15$ mK in amplitude, as in the left panel of Figure \ref{fig:gs_Z}. 

\end{document}